\documentclass[journal]{IEEEtran}
\ifCLASSINFOpdf
\else
\fi
\usepackage{cite}
\usepackage{color}
\usepackage{graphicx}
\usepackage{lineno}
\usepackage{amsmath}
\usepackage{amssymb,amsfonts}
\usepackage{booktabs}
\usepackage{threeparttable}
\usepackage{subfig}
\usepackage{multirow}

\begin{document}
%
\title{MLBF-Net: A Multi-Lead-Branch Fusion Network for Multi-Class Arrhythmia Classification Using 12-Lead ECG}
%
%
%

\author{Jing Zhang, Deng Liang, Aiping Liu, Min Gao, Xiang Chen, Xu Zhang, Xun Chen
\thanks{Xun Chen is the \textit{Corresponding author}.}
\thanks{Jing Zhang, Deng Liang, Aiping Liu, Xiang Chen, Xu Zhang, Xun Chen are with the School of Information Science and Technology, University of Science and Technology of China, Hefei 230027, China. Min Gao is with the Department of Electrocardiogram, The First Affiliated Hospital of University of Science and Technology of China (Anhui Provincial Hospital), Hefei 230036, China.
(e-mail: zjing5@mail.ustc.edu.cn, ld1995@mail.ustc.edu.cn, aipingl@ustc.edu.cn, Gmbeauty@163.com, xch@ustc.edu.cn, xuzhang90@ustc.edu.cn, xunchen@ustc.edu.cn)}}

%
%



\maketitle

\begin{abstract}
  Automatic arrhythmia detection using 12-lead electrocardiogram (ECG) signal plays a critical role in early prevention and diagnosis of cardiovascular diseases. In the previous studies on automatic arrhythmia detection, most methods concatenated 12 leads of ECG into a matrix, and then input the matrix to a variety of feature extractors or deep neural networks for extracting useful information. Under such frameworks, these methods had the ability to extract comprehensive features (known as integrity) of 12-lead ECG since the information of each lead interacts with each other during training. However, the diverse lead-specific features (known as diversity) among 12 leads were neglected, causing inadequate information learning for 12-lead ECG. To maximize the information learning of multi-lead ECG, the information fusion of comprehensive features with integrity and lead-specific features with diversity should be taken into account. In this paper, we propose a novel Multi-Lead-Branch Fusion Network (MLBF-Net) architecture for arrhythmia classification by integrating multi-loss optimization to jointly learning diversity and integrity of multi-lead ECG. MLBF-Net is composed of three components: 1) multiple lead-specific branches for learning the diversity of multi-lead ECG; 2) cross-lead features fusion by concatenating the output feature maps of all branches for learning the integrity of multi-lead ECG; 3) multi-loss co-optimization for all the individual branches and the concatenated network. We demonstrate our MLBF-Net on China Physiological Signal Challenge 2018 which is an open 12-lead ECG dataset. The experimental results show that MLBF-Net obtains an average $F_1$ score of 0.855, reaching the highest arrhythmia classification performance. The proposed method provides a promising solution for multi-lead ECG analysis from an information fusion perspective.
\end{abstract}


%
\IEEEpeerreviewmaketitle

\section{Introduction}

\IEEEPARstart{C}{ardiovascular} disease (CVD) is the leading cause of global mortality. It was announced by the World Health Organization (WHO) that an estimated 17.9 million people died from CVD in 2016, accounting for 31\% of global deaths \cite{CVDs}. Cardiac arrhythmia is a very common type of CVD, which manifests as abnormal heart rhythms. According to statistics, about half of all cardiovascular deaths are sudden cardiac deaths and about 80\% of these are caused by cardiac arrhythmia \cite{mehra2007global}. Electrocardiogram (ECG) is a widely accessed, non-invasive, and inexpensive tool for arrhythmia diagnosis in clinic. It records the heart's electrical activities over time through electrodes attached to the skin surface. Recently, intelligent healthcare has become increasingly prominent. Automatic arrhythmia detection based on ECG could assist doctors in clinical practice, and also provide ordinary people with daily monitoring using wearable devices. Therefore, how to promote the accuracy of automatic arrhythmia detection is a critical issue.

Over the last decades, a large number of traditional ECG classification methods have been developed. Traditional methods primarily comprise three procedures involving preprocessing (e.g. denoising and heartbeat segmentation), feature extraction, and classification. Of these, feature extraction is the most crucial step, which relies on professional knowledge to construct a set of hand-craft features. In recent years, deep learning has made great success in the field of healthcare with its powerful capability to extract high-level abstract features automatically, avoiding laborious manual feature design. Many studies have designed deep learning-based approaches for arrhythmia detection using ECG signals.

A standard ECG record contains 12 leads (i.e. I, II, III, avR, avL, avF, V1, V2, V3, V4, V5, V6), which is widely used in clinical arrhythmia diagnosis. The 12-lead ECG has two inherent properties: integrity and diversity \cite{liu2018multiple-feature-branch}. On the one hand, the 12-lead ECG signal contains comprehensive information by recording the electrical potential from different spatial angles of the heart. It gives an overall reflection on the heart's condition. Thus, the 12-lead ECG could be treated as an integrated one to make a diagnosis. On the other hand, different leads correspond to different anatomical areas of the heart, providing distinct perspectives. Thus, the ECG signal under each lead contains lead-specific features, and the 12-lead ECG signal has diverse information across leads. To maximize information learning for 12-lead ECG, integrity and diversity should be both taken into account. For multi-lead ECG analysis, most existing studies concatenated multi-lead ECG into a matrix, and then input the matrix to a variety of feature extractors or deep neural networks for extracting useful information. However, such approaches lacked explicit mechanisms to realize lead-specific features extraction and only considered comprehensive features extraction of multi-lead ECG. Furthermore, they were difficult to utilize the adequate information fusion of diversity and integrity to enhance the detecting performance.

To address the limitations mentioned above, how to fully utilize the diversity and integrity of multi-lead ECG and thereby maximize the information learning is investigated in this paper. The main contributions of this paper are concluded below:

1) We propose a novel Multi-Lead-Branch Fusion Network (MLBF-Net) architecture with multiple branches for arrhythmia classification using 12-lead ECG. For realizing diversity learning, each branch of MLBF-Net is designed to classify the ECG signal under a specific lead, which could learn lead-specific features. Specifically, each branch introduces a hierarchical network structure consisting of the convolutional layers, bidirectional gated recurrent unit (BiGRU), and an attention module to mine the discriminative information further. For realizing integrity learning, the output feature maps from all branches are concatenated to form the concatenated network of MLBF-Net. It is responsible for learning to classify the ECG signal based on all 12 leads, which could extract comprehensive features.

2) We design a collaborative optimization strategy with multiple losses, specialized for all the individual branches and the concatenated network. This strategy not only optimizes comprehensive features of multi-lead ECG, but also realizes lead-specific features learning simultaneously during the training process, which could achieve the information fusion of diversity and integrity.

The remainder of this paper is arranged as follows. Section II outlines the related works. Section III describes the architecture of the proposed Multi-Lead-Branch Network. Section IV presents the experimental result. Section V gives the discussions. Finally, Section VI summaries this paper.

\section{Related Works}
Traditional ECG classification methods designed a number of hand-craft features. Typical hand-craft features include statistical features \cite{li2013ventricular,dilmac2015ecg,hurnanen2016automated}, P-QRS-T features \cite{tsipouras2005arrhythmia,haseena2011fuzzy,mar2011optimization}, morphological features \cite{mar2011optimization,kutlu2011multi,rodriguez2009unsupervised,zhang2014heartbeat}, and wavelet features \cite{ozbay2009new,khorrami2010comparative,seera2015classification,elhaj2016arrhythmia}. Also, mathematical transformations that transform the high-dimensional ECG signal into a lower-dimensional space can be used for extracting meaningful information, such as independent component analysis (ICA) \cite{martis2013application,ye2015automatic,martis2013ecg}, principal component analysis (PCA) \cite{ince2009generic,martis2013ecg,wang2013ecg}, and linear discriminant analysis (LDA) \cite{martis2013ecg,wang2013ecg}. Following feature extraction, a variety of classifiers are carried out to classify the extracted features. This can be implemented by artificial neural network (ANN) \cite{ozbay2009new,haseena2011fuzzy}, support vector machine (SVM) \cite{elhaj2016arrhythmia,zhang2014heartbeat,li2013ventricular,khorrami2010comparative}, k nearest neighbor (KNN) \cite{martis2013application,kutlu2011multi}, decision tree \cite{martis2013application,seera2015classification} and bayesian classifier \cite{zhang2014heartbeat,mar2011optimization}.

Deep learning is increasingly predominant in recent studies on ECG classification. Convolutional neural networks (CNNs) are a commonly adopted type of deep neural network due to its effectiveness in extracting features. Kiranyaz et al. \cite{kiranyaz2015real} designed an adaptive CNN for patient-specific ECG heartbeat classification, which incorporates traditional feature extraction and classification into a single learning structure. Rahhal et al. \cite{al2018convolutional} transformed ECG signals to image-like representations using continuous wavelet transform, and then fed these representations into a deep CNN pretrained on ImageNet with a large number of annotated images, which achieved a good detection performance for supraventricular ectopic beats and ventricular ectopic beats. Rajpurkar et al. \cite{hannun2019cardiologist} presented a 34-layer residual CNN with a cardiologist-level accuracy in detecting twelve cardiac arrhythmias. In other studies, the ECG signal was viewed as a time-series and recurrent neural network (RNN) specialized for tackling sequential data was adopted. The representative variants of RNN include long short-term memory (LSTM) and gated recurrent unit (GRU). Saadatnejad et al. \cite{saadatnejad2019lstm} developed a real-time heartbeat classification algorithm for personal wearable devices based on multiple LSTMs and wavelet transform. Lynn et al. \cite{lynn2019deep} proposed a deep bidirectional GRU network for classifying biometric ECG signals. Further, many research works have designed hierarchical networks by stacking CNN and RNN. He et al. \cite{he2019automatic} stacked a deep residual CNN and a bidirectional LSTM layer for arrhythmia classification, and obtained a good performance. Yao et al. \cite{yao2018time} classified multi-class arrhythmias using an integrated model consisting of VGGNet-based CNN and varied-length LSTMs, which allows for varied-length signal input and is effective in detecting paroxysmal arrhythmias.

Recently, Liu et al. \cite{liu2018multiple-feature-branch} proposed a multiple-feature-branch convolutional neural network (MFB-CNN) for myocardial infarction detection using 12-lead ECG, aiming at exploiting both integrity and diversity. Particularly, 12 ECG leads were input into different branches of MFB-CNN for learning lead-specific features; a fully connected layer was used to summarize the output feature map of 12 feature branches, utilizing the integrity. However, the parameters of MFB-CNN were optimized by only a single loss function in an end-to-end way. This single-loss training strategy is unable to fully exploit the specific information of each individual branch in an isolated way, since the classification loss of training samples is backwardly propagated to all 12 branches \cite{qi2019learning}. Therefore, diversity learning is weakened significantly. In \cite{liu2020mfb-cbrnn:}, Liu et al. refined the network architecture of MFB-CNN, in which the fully connected layer was replaced by LSTM for summarizing all the branches. In addition, several branches were randomly deactivated at each training iteration for improving the generalization of the model. Nevertheless, as with \cite{liu2018multiple-feature-branch}, the deficiency in diversity learning still exists. Currently, the existing works have not achieved joint learning for the diversity and integrity of multi-lead ECG. In other words, the fusion of diversity and integrity to maximize the information learning of multi-lead ECG is worthy of further investigation.

\section{Methods}
\subsection{Model Overview}
\begin{figure*}[!t]
\centering
\includegraphics[width=0.9\textwidth]{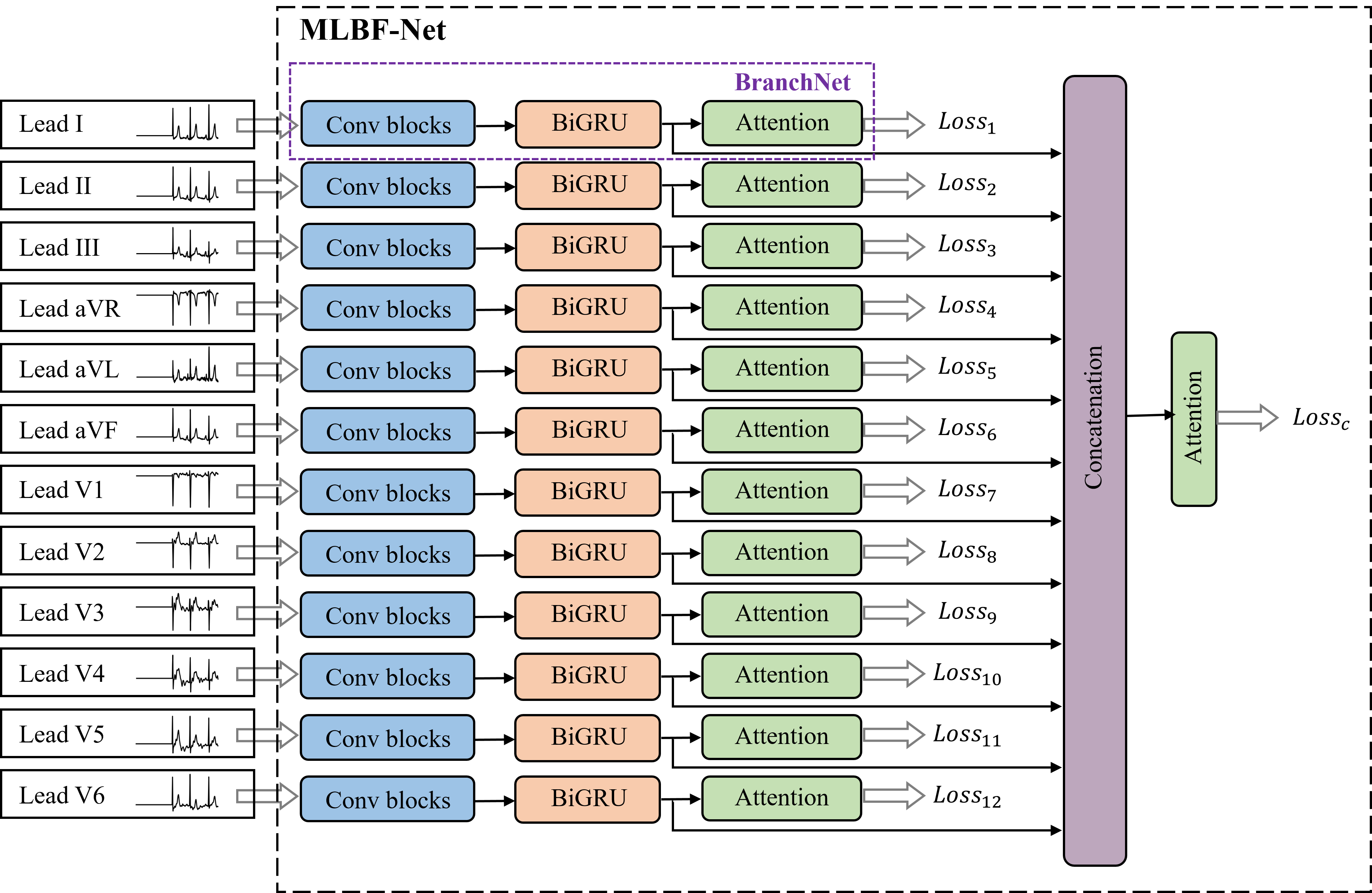}
\caption{The proposed network architecture.}
\label{model}
\end{figure*}

The proposed Multi-Lead-Branch Fusion Network (MLBF-Net) is illustrated in Fig. \ref{model}. It is mainly composed of three components: 1) multiple lead-specific branches for learning the diversity of multi-lead ECG; 2) cross-lead features fusion by concatenating the output feature maps of all branches for learning the integrity of multi-lead ECG; 3) multi-loss co-optimization for all the individual branches and the concatenated network. These three components are described in detail below.

\subsection{Single Lead-Branch Feature Learning}
\begin{table}[!t]
  \caption{\label{BranchNet}The configuration of each branch.}
  \centering
  \begin{tabular}{c}
    \centering
	\includegraphics[width=0.9\linewidth]{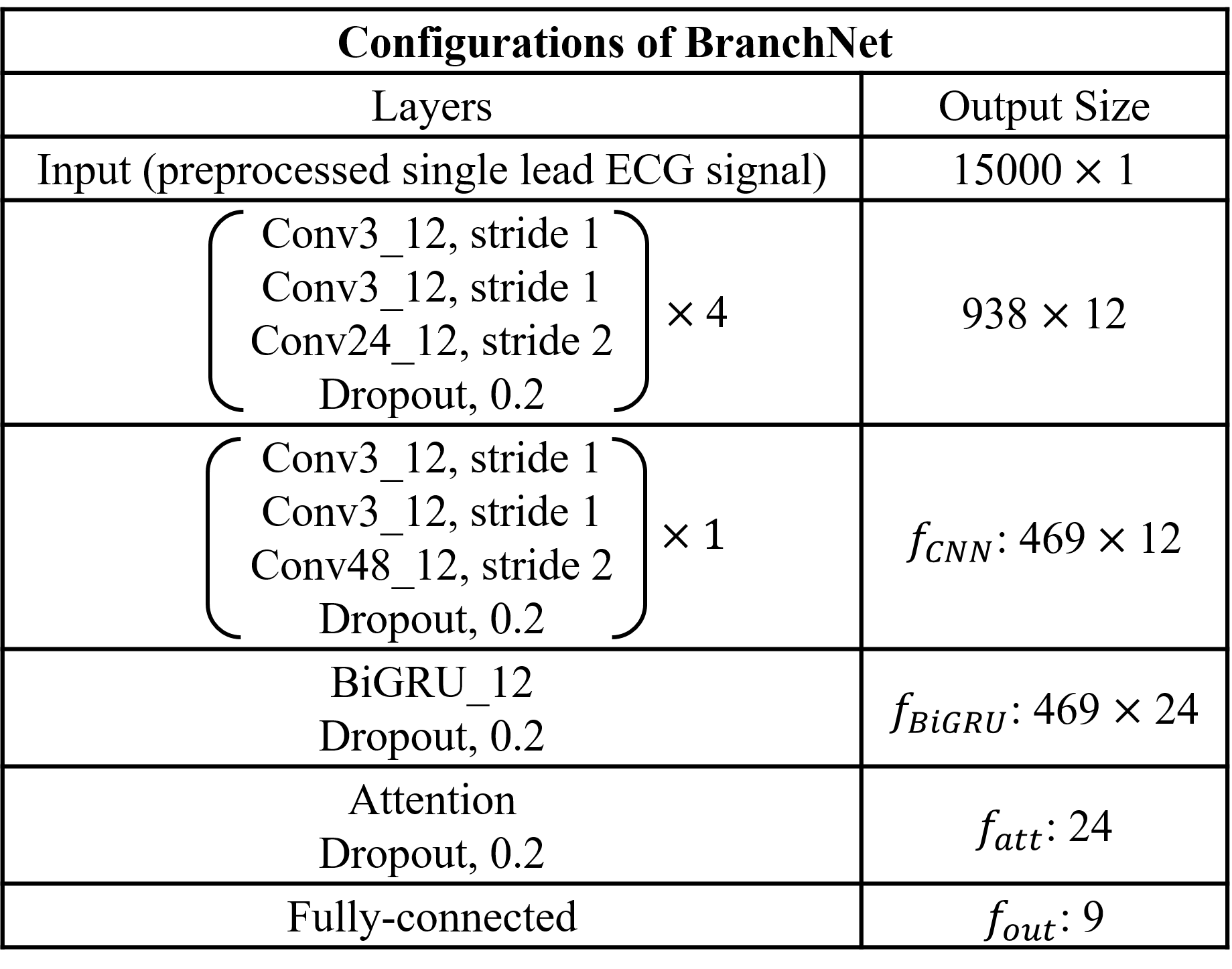}
  \end{tabular}
  \begin{tablenotes}{\footnotesize
  \item[1] The convolutional layer is represented as Conv(kernel size)\_(kernel number).}
  \end{tablenotes}
\end{table}

The preprocessed 12-lead ECG signals $X$ are split into 12 single leads, denoted as $X_j \in \mathbb{R}^{L \times 1}$, $j \in \{1,2,...,12\}$ where L is the length of the preprocessed signal. $X_j$ is then fed into the jth branch. The configuration of all branches is the same. For convenience, we name the single branch network "BranchNet", of which the configuration is shown in Table \ref{BranchNet}. BranchNet combines 15 convolutional layers, a bidirectional gated recurrent unit (BiGRU) layer, and an attention module. Its internal behavior can be formulated as:
\begin{gather}
f_{{cnn}_j} = CNN(X_j) \\
f_{{BiGRU}_j} = BiGRU(f_{{cnn}_j}) \\
f_{{att}_j} = Attention(f_{{BiGRU}_j})
\end{gather}
where $CNN$, $BiGRU$, and $Attention$ denote the convolutional neural subnetwork, BiGRU layer, and attention module, respectively. $f_{{cnn}_j}$, $f_{{BiGRU}_j}$, and $f_{{att}_j}$ denote the output feature map of these three modules of the jth branch.

The cross-entropy loss is employed for training single branch, calculated as:
\begin{equation}
L_j = -\frac{1}{N}\sum\limits_{i=1}^N{\rm log}(\frac{{\rm exp}(p(X_j^{(i)},y^{(i)}))}{\sum\nolimits_c {\rm exp}(p(X_j^{(i)},c))})
\end{equation}
where N is the number of training samples, $y^{(i)}$ is the true label of ith sample, and $p(X_j^{(i)},c)$ denotes the probability that the input $X_j^{(i)}$ is predicted as label $c$. By minimizing the cross-entropy loss function during training, lead-specific features are iteratively optimized for achieving diversity learning.

\textbf{Convolutional Neural Subnetwork:} As shown in Table \ref{BranchNet}, the convolutional neural subnetwork is composed of 5 convolutional blocks, with a total of 15 convolutional layers. Each convolution block includes three convolution layers, together with a dropout \cite{srivastava2014dropout} layer. The output of each convolutional layer is nonlinear transformed by a leaky rectified linear unit (LeakyReLU) activation function, where the operation is omitted for brevity in Table \ref{BranchNet}. Although ReLU \cite{nair2010rectified} is a more common choice, LeakyReLU \cite{maas2013rectifier} is applied due to the ability to avoid the dead neurons problem. To mitigate the neural network from overfitting, the dropout rate is set to 0.2.

\begin{figure}[tbp]
\centering
\includegraphics[width=0.45\textwidth]{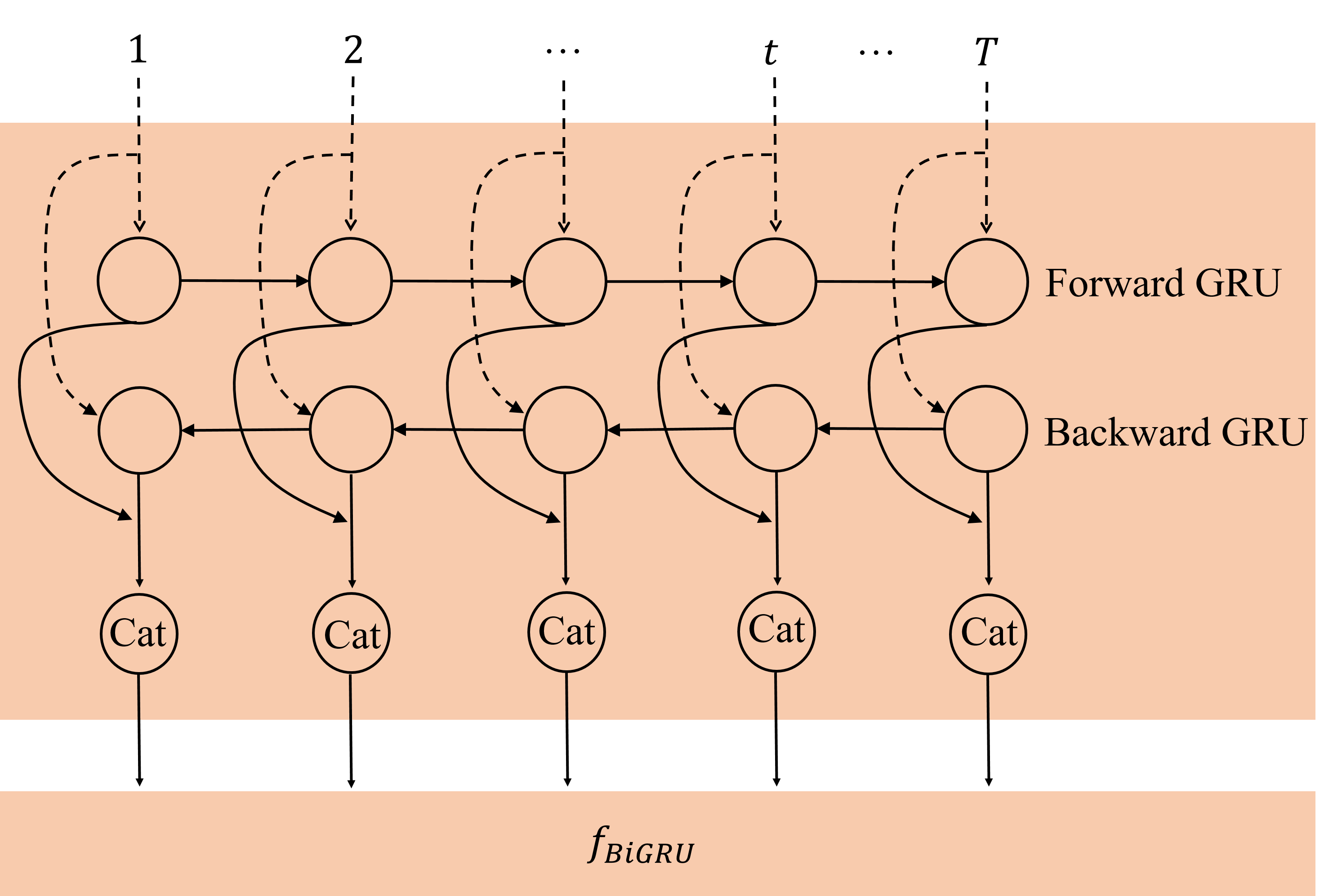}
\caption{The illustration for a bidirectional GRU. "Cat" is an abbreviation for concatenation operation.}
\label{BiGRU}
\end{figure}

\textbf{Bidirectional GRU:} The output feature map $f_{cnn}$ of convolutional neural subnetwork flows into a bidirectional GRU layer. GRU \cite{cho-etal-2014-learning} and LSTM \cite{hochreiter1997long} are the evolutionary implementations of RNN. We select GRU since it has similar performance to LSTM but with less computational complexity. As illustrated in Fig. \ref{BiGRU}, the bidirectional GRU consists of a forward GRU and a backward GRU, which read the time-series feature map in the temporal and anti-temporal directions, respectively. At time $t$, the forward GRU aggregates the information of $f_{cnn}$ from 1 to $t$ and the backward GRU aggregates the information of $f_{cnn}$ from T to $t$. In our model, T equals 469, which is the time length of $f_{cnn}$. The features from both opposite directions are incorporated by the bidirectional GRU to obtain the contextual information. Note that we achieve the incorporation by concatenation operation, summarizing the features centered around time $t$. In our experiments, the unit number of the bidirectional GRU layer is configured to 12, meaning that the dimension for each time step is 12.

\textbf{Attention Module:} Different ECG components such as P wave, QRS complex wave, and ST-segment contribute differentially to the pathological information of the whole ECG segment. Hence, attention mechanism \cite{yang2016hierarchical} is introduced to extract the informative segments and merge the representation of these important segments. This process is formulated as:
\begin{gather}
u_{jt} = tanh(W_wf_{BiGRU_{jt}} + b_w) \\
\alpha_{jt} = \frac{exp(u^T_{jt}u_w)}{\sum_t exp(u^T_{jt}u_w)} \\
f_{att_j} = \sum_t \alpha_{jt}f_{BiGRU_{jt}}
\end{gather}
That is, the encoded ECG signal $f_{BiGRU_{jt}}$ is fed into a one-layer multilayer perceptron to obtain $u_{jt}$ as a hidden representation of $f_{BiGRU_{jt}}$. Then the similarity of $u_{jt}$ and a trainable vector $u_w$ is measured and a normalized importance weight vector $\alpha_{jt}$ is obtained through a softmax function. Afterwards, the weighted sum of the encoded ECG signal $f_{BiGRU_{jt}}$ and its corresponding weight vector $\alpha_{jt}$ is computed to get the weighted representation $f_{{att}_j}$. $W_w$, $u_w$, and $b_w$ are randomly initialized trainable parameters.

\subsection{Multi-Lead Feature Fusion Learning}
For learning complementary cross-lead information and providing more robust diagnosis, we fuse the features extracted from each lead by concatenating the lead-specific feature maps $f_{{BiGRU}_j}$ of all branches in channel axis:
\begin{equation}
F = Cat(f_{{BiGRU}_1}, f_{{BiGRU}_2}, ..., f_{{BiGRU}_{12}})
\end{equation}
Then the concatenated feature map $F$ is fed through the same attention module used in BranchNet to build the concatenated network. The concatenated network shares the feature maps from the input $X_j$ to the output $f_{{BiGRU}_j}$ of BiGRU layer with each branch for obtaining the comprehensive prediction based on 12-lead ECG. Similar to single lead-branch training, a cross-entropy loss is also employed to train the concatenated network, calculated as:
\begin{equation}
L_c = -\frac{1}{N}\sum\limits_{i=1}^N{\rm log}(\frac{{\rm exp}(p(X^{(i)},y^{(i)}))}{\sum\nolimits_c {\rm exp}(p(X^{(i)},c))}) \\
\end{equation}
where $p(X^{(i)},c)$ denotes the probability that the input $X^{(i)}$ is predicted as label $c$.

\begin{figure*}[!t]
\centering
\includegraphics[width=0.8\textwidth]{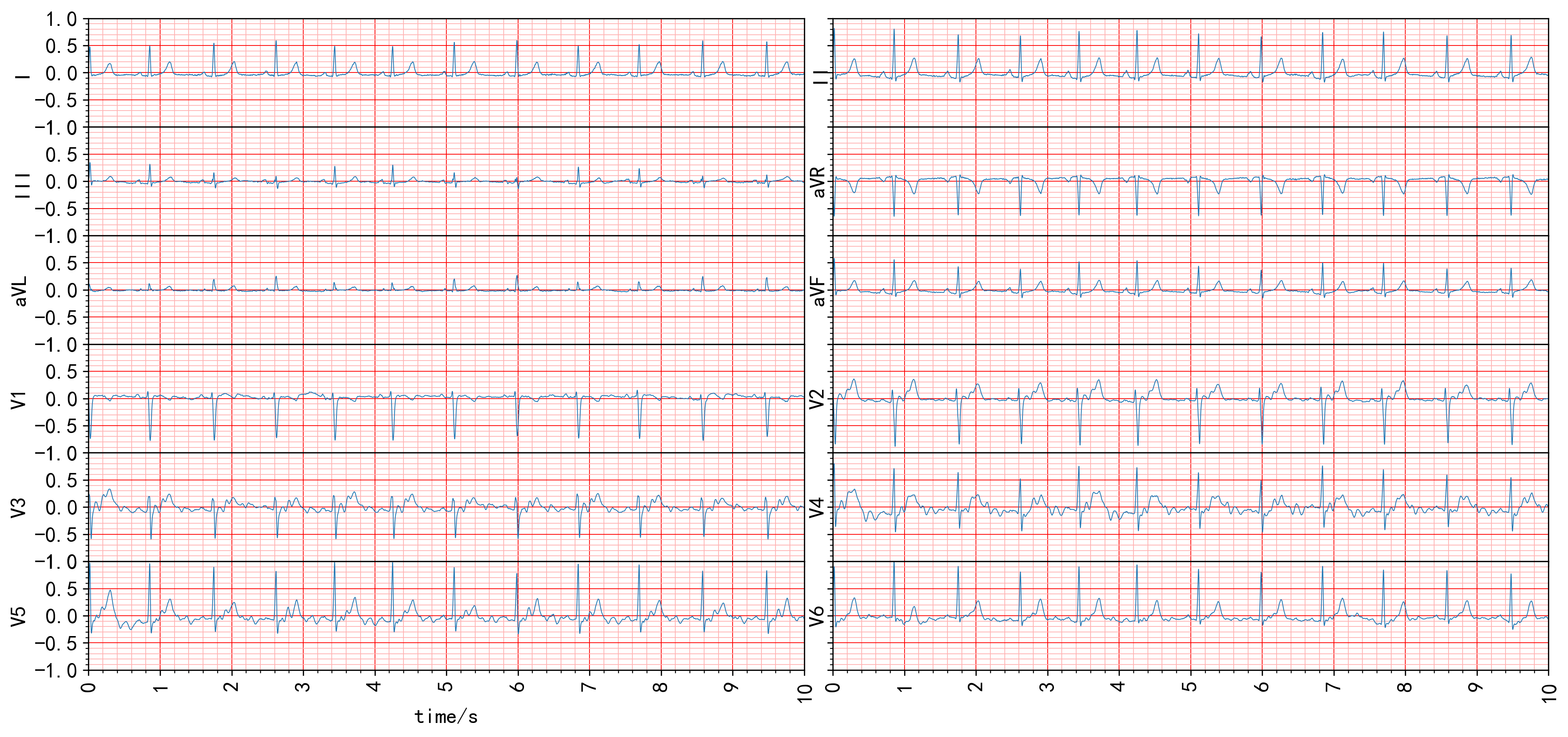}
\caption{An example of normal 12-lead ECG record.
\label{ecg}}
\end{figure*}

\subsection{Joint Optimization with Multiple Losses}
It is noted that lead-specific branches are interrelated rather than independent. In order to jointly learning the diversity and integrity of multi-lead ECG, we design multiple dedicated losses for collaboratively optimizing multiple branches and the concatenated network. The loss for each branch is designed to optimize lead-specific features for maximizing the discriminative capability of single-lead ECG, and the loss for the concatenated network aims to optimize multi-lead comprehensive features simultaneously. For the training of the whole model, the final loss function is defined as:
\begin{equation}
L = L_c + \lambda(L_1 + ... + L_{12})
\end{equation}
where $\lambda$ is a balance parameter used for determining the importance ratio between diversity and integrity, which is set to 1 in our experiments. $L_c$ and $L_j$ are the cross-entropy losses of the concatenated network and the jth branch, respectively.

The co-optimization strategy not only optimizes multi-lead comprehensive features, but also realizes lead-specific features learning simultaneously during the training process. Compared with the regular single-loss learning, this strategy achieves the information fusion of diversity and integrity, thereby promoting the maximum of multi-lead ECG information learning.

\section{Experiment}
The proposed method is performed using Python language and Keras 2.2.4 framework. All experiments in this paper were run on a server with Xeon E5 2620 CPU, 128GB memory and four GeForce RTX cards.

\subsection{Data Description}
China Physiological Signal Challenge 2018 (CPSC 2018) provides 12-lead ECG records, which is suitable for multi-lead ECG analysis. The proposed network was trained using the public dataset of CPSC 2018, and evaluated on the private test set that is inaccessible for researchers to ensure a fair comparison. The ECG records were acquired from 11 hospitals. The sampling rate is 500Hz. The public dataset includes 6877 12-lead ECG records varying in length from 6 to 60 seconds, and the test set includes 2954 12-lead ECG records of similar lengths. The details of the public database can be found in Table \ref{dataset}, with a total of 9 ECG classes containing normal rhythm and 8 types of arrhythmias. Figure \ref{ecg} gives an example of a 12-lead ECG record. More details about CPSC 2018 database can be seen in \cite{liu2018open}.

\begin{table}[!t]
  \centering
  \caption{\label{dataset}Distribution of ECG classes on the public dataset of CPSC 2018.}
  \begin{tabular}{c}
    \centering
	\includegraphics[width=0.9\linewidth]{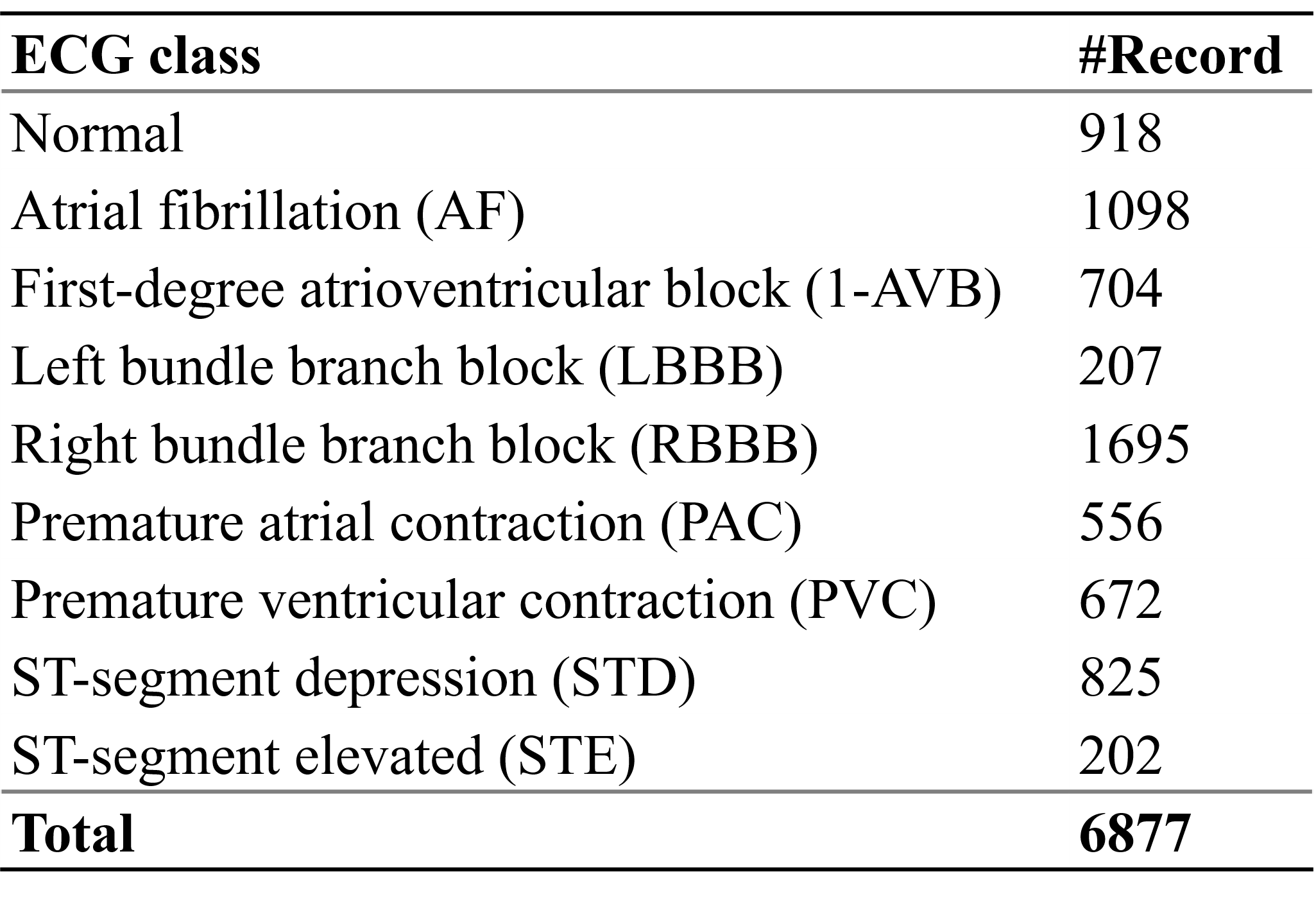}
  \end{tabular}
\end{table}

\subsection{Evaluation Metric}
In this research, $F_1$ score is used to measure the model's classification performance for each class. It is the harmonic mean of precision and recall, defined as:
\begin{gather}
Precision = \frac{TP}{TP + FP} \\
Recall = \frac{TP}{TP + FN} \\
F_1 = \frac{2 \times (Precision \times Recall)}{Precision+Recall}
\end{gather}
The terms TP, FP, and FN refer to the sample number of true positive, false positive, and false negative, respectively. The average of $F_1$ scores for all classes, namely macro-$F_1$ score, is computed to make a final evaluation.

\subsection{Implementation Details}
\newcommand{\tabincell}[2]{\begin{tabular}{@{}#1@{}}#2\end{tabular}}
\begin{table*}[!t]
\centering
\caption{\label{comparison}Comparison for classification performance of previous works and ours evaluated on the private test set of CPSC 2018.}
  \resizebox{\textwidth}{!}{
  \begin{tabular}{lllllllllllll}
    \toprule
    \multirow{2}{*}{Work} &\multirow{2}{*}{Year} &\multirow{2}{*}{Architecture} &\multicolumn{10}{c}{$F_1$ score} \\
    \cmidrule{4-13}
    &&&\tabincell{l}{Normal\\(394)} &\tabincell{l}{AF\\(466)} &\tabincell{l}{I-AVB\\(295)} &\tabincell{l}{LBBB\\(97)} &\tabincell{l}{RBBB\\(756)} &\tabincell{l}{PAC\\(250)} &\tabincell{l}{PVC\\(276)} &\tabincell{l}{STD\\(340)} &\tabincell{l}{STE\\(80)} &\tabincell{l}{Average\\(2954)} \\
    \midrule
    \cite{yao2018time}      &2018 &CNN+LSTM                              &0.753 &0.900 &0.809 &0.874 &0.922 &0.638 &0.832 &0.762 &0.462 &0.772 \\
    \cite{liu2018automatic} &2018 &Expert features; CNN                  &0.82  &0.91  &0.87  &0.87  &0.91  &0.63  &0.82  &0.81  &0.60  &0.81  \\
    \cite{he2019automatic}  &2019 &CNN+LSTM                              &- &- &- &- &- &- &- &- &-                                     &0.806 \\
    \cite{wang2019multi}    &2019 &CNN+Attention                         &0.79  &0.93  &0.85  &0.86  &0.93  &0.75  &0.85  &0.80  &0.56  &0.813 \\
    \cite{yao2020multi}     &2020 &CNN+LSTM+Attention                    &0.789 &0.920 &0.850 &0.872 &0.933 &0.736 &0.861 &0.789 &0.556 &0.812 \\
    \cite{9064540}          &2020 &Multi-Scaled CNN+Attention            &0.82  &0.90  &0.86  &0.87  &0.93  &0.78  &\textbf{0.88}  &0.80  &0.62  &0.828 \\
    \cite{ZHANG2020101856}  &2020 &CNN+Attention+BiGRU; Ensemble model   &0.819 &\textbf{0.936} &0.866 &0.862 &0.926 &0.789 &0.865 &0.812 &0.640 &0.835 \\
    \cite{chen2020detection}&2020 &CNN+BiGRU+Attention; Ensemble model   &0.801 &0.933 &0.875 &0.884 &0.910 &\textbf{0.826} &0.869 &0.811 &0.624 &0.837 \\
    Ours$^1$                &     &Multi-Lead-Branch; CNN+BiGRU+Attention&\textbf{0.850} &0.933 &\textbf{0.885} &0.862 &0.931 &0.801 &0.859 &\textbf{0.820} &0.646 &0.843 \\
    Ours$^2$          &&Ensemble model based on Ours$^1$ &0.847 &0.934 &0.884 &\textbf{0.896} &\textbf{0.939} &0.822 &0.878 &0.818 &\textbf{0.677} &\textbf{0.855} \\
    \bottomrule
  \end{tabular}}
  \begin{tablenotes}{\footnotesize
  \item[1] The "-" indicates the $F_1$ score was not reported. The number of ECG classes are in parentheses.
  \item[2] Ours$^1$ indicates the single model without ensemble. Ours$^2$ indicates ensemble model.
  \item[3] The highest score for each class is in bold.}
  \end{tablenotes}
\end{table*}

\begin{table*}[!t]
\centering
  \caption{\label{table3}Classification performance comparison for single-branch and multi-branch frameworks with 12-lead ECG as input by 10-fold cross validation (mean$\pm$SD).}
  \resizebox{\textwidth}{!}{
  \begin{tabular}{ccccccccccc}
    \toprule
    Network & Normal & AF & I-AVB & LBBB & RBBB & PAC & PVC & STD & STE & average $F_1$ \\
    \midrule
    BranchNet & 0.809$\pm$0.018 & 0.914$\pm$0.017 & 0.872$\pm$0.022 & 0.882$\pm$0.043 & 0.929$\pm$0.019 & 0.739$\pm$0.060 & 0.872$\pm$0.027 & 0.808$\pm$0.033 & 0.513$\pm$0.101 & 0.815$\pm$0.018 \\
    MLBF-Net & \textbf{0.832$\pm$0.033} & \textbf{0.932$\pm$0.017} & \textbf{0.902$\pm$0.025} & \textbf{0.911$\pm$0.032} & \textbf{0.944$\pm$0.013} & \textbf{0.814$\pm$0.051} & \textbf{0.889$\pm$0.018} & \textbf{0.834$\pm$0.038} & \textbf{0.608$\pm$0.126} & \textbf{0.852$\pm$0.021} \\
    \bottomrule
  \end{tabular}}
\end{table*}

\subsubsection{Preprocessing}
Similar to our previous work \cite{ZHANG2020101856}, the preprocessing for the original ECG signal includes two procedures: downsampling and cropping or padding. In the first step, the downsampling from 500Hz to 250Hz was performed to speed up the training. In the second step, the downsampled ECG signals were cropped or padded with zeros to the same length because convolutional neural networks do not accept varied-length input. In our setting, 60 seconds was the target length. It means that the signals longer than 60 seconds were cropped and those less than 60 seconds were padded with zero.

\subsubsection{Training Setting}
We trained the proposed model in an end-to-end way. The preprocessed training data was grouped into batches of 64 samples to fed into MLBF-Net. We set 64 to batch size by tuning this hyperparameter. The batched updating of network parameters takes less memory and is more computationally efficient. More importantly, a more robust convergence can be obtained by mini-batch, averting local minima. We applied adaptive moment estimation (Adam) optimizer \cite{kingma2014adam} to update the weights of the whole model iteratively with a fixed learning rate of 0.001. Adam, which updates weights based on exponential decaying averages of past gradients and past squared gradients, is usually considered to converge to an excellent performance \cite{fan2018multiscaled}.

In order to alleviate the proposed model overfitting during the training process, another two strategies were adopted in addition to dropout layers of the network structure. The first strategy is earlystopping, which stops training if the classification performance of the model on validation data remains unimproved up to 50 epochs. The best-performing model was saved. The second strategy is that we set macro-$F_1$ score of validation data as the stopping criterion. Different from the accuracy metric that is dominated by the classes with more samples, macro-$F_1$ score is an unbiased metric towards unbalanced ECG classes.

\subsection{Classification Performance}
Table \ref{comparison} shows the classification performance of the proposed method, and makes a comparison with eight previous works in detail. To ensure a fair comparison, our models were evaluated on the private test set of CPSC 2018 that is also used in the evaluation of these works. In order to provide a more robust prediction, We applied 10-fold cross validation to get an ensemble model, denoted as Ours$^2$. In particular, the public dataset was randomly split into ten subsets, with each subset in turn as validation data and the rest ones as training data. Ten training data sets were separately preprocessed by the above preprocessing operations and then input into the proposed network architecture to obtain ten models. Finally, the prediction probabilities of these models were averaged as the final probabilities. In addition, a single model without ensemble was also evaluated, denoted as Ours$^1$.

\cite{chen2020detection} was the first place among CPSC 2018 competitors, where their average $F_1$ score is 0.837. As shown in Table \ref{comparison}, the proposed method outperforms the existing methods in the average screening capability for 9 types of ECG. It even beats the previous first place by 1.8\% average $F_1$ score, achieving the highest classification performance. It is observed that our model makes a superior diagnosis than other models for Normal, I-AVB, LBBB, RBBB, STD, and STE. Among them, the most significant superiority lies in the identification of Normal and STE, gaining 2.7\% (0.82-0.847), 3.7\% (0.64-0.677) $F_1$ score increase than the previous best-performing ones, respectively.

\section{Discussion}
\begin{figure*}[!t]
\centering
\subfloat[The performance visualization of MLBF-Net]{
\label{multi-lead}
\includegraphics[width=0.29\textwidth]{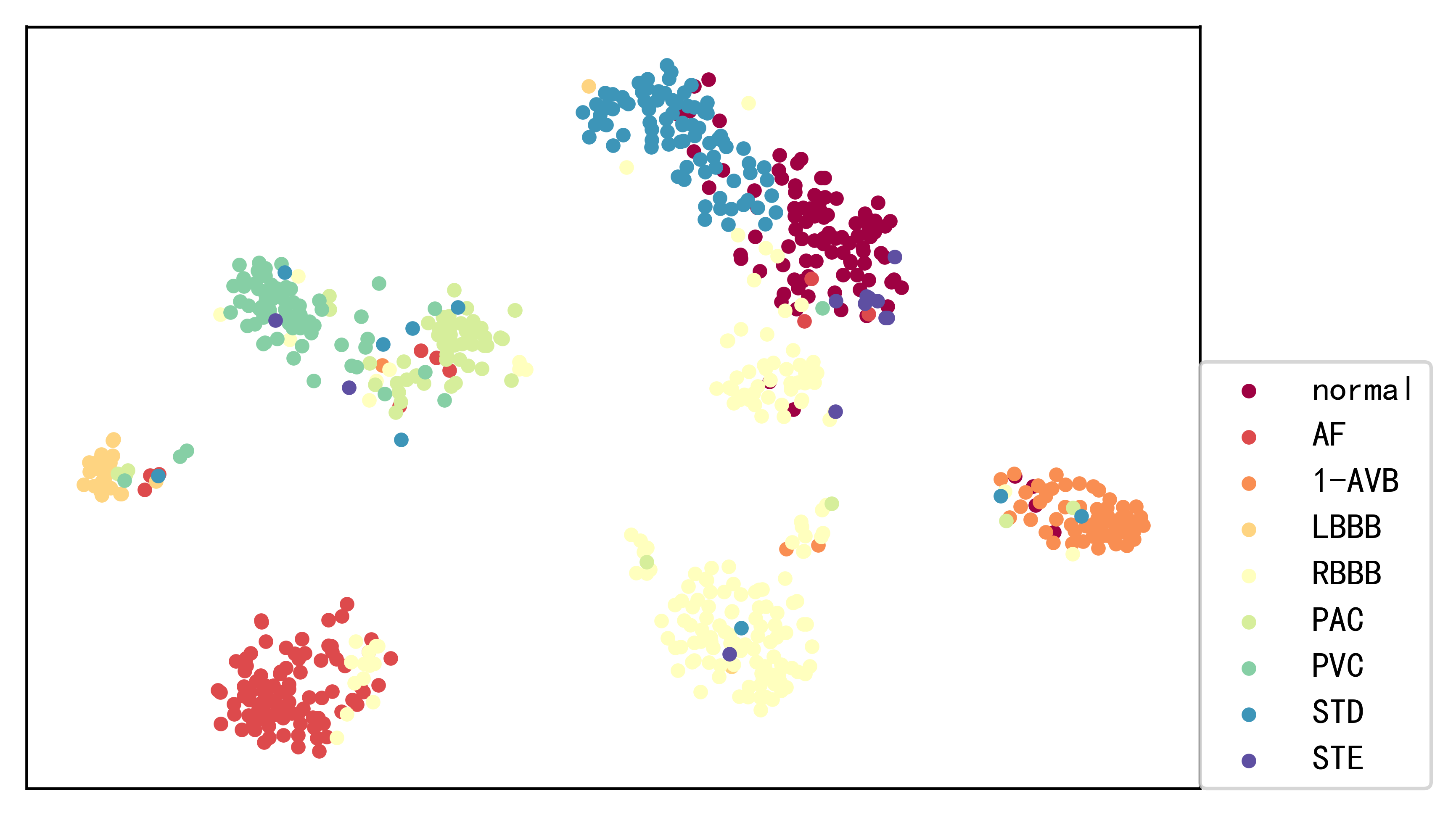}} \\
\subfloat[branch for lead I]{
\label{I}
\includegraphics[width=0.29\textwidth]{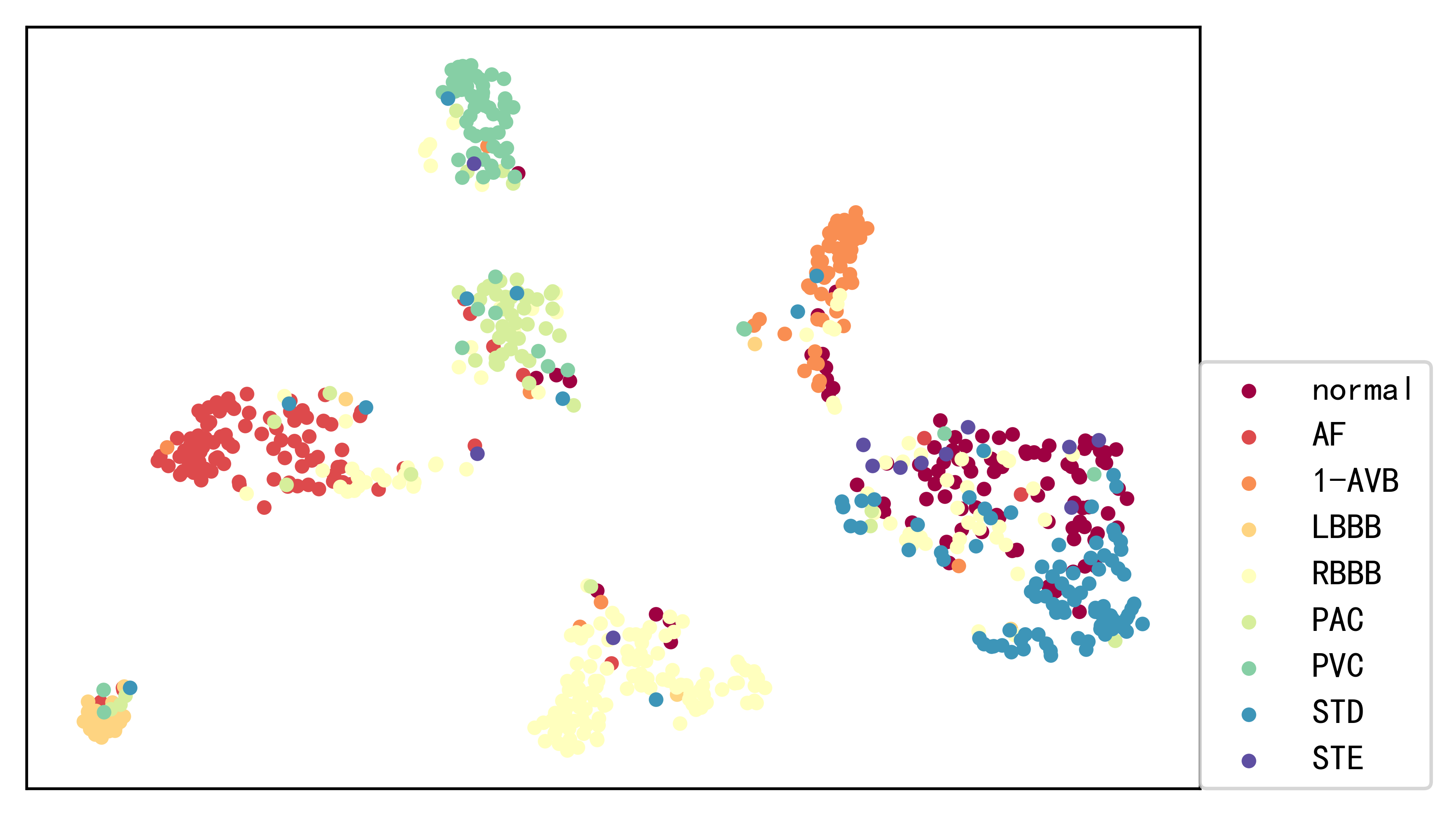}}
\subfloat[branch for lead II]{
\label{II}
\includegraphics[width=0.29\textwidth]{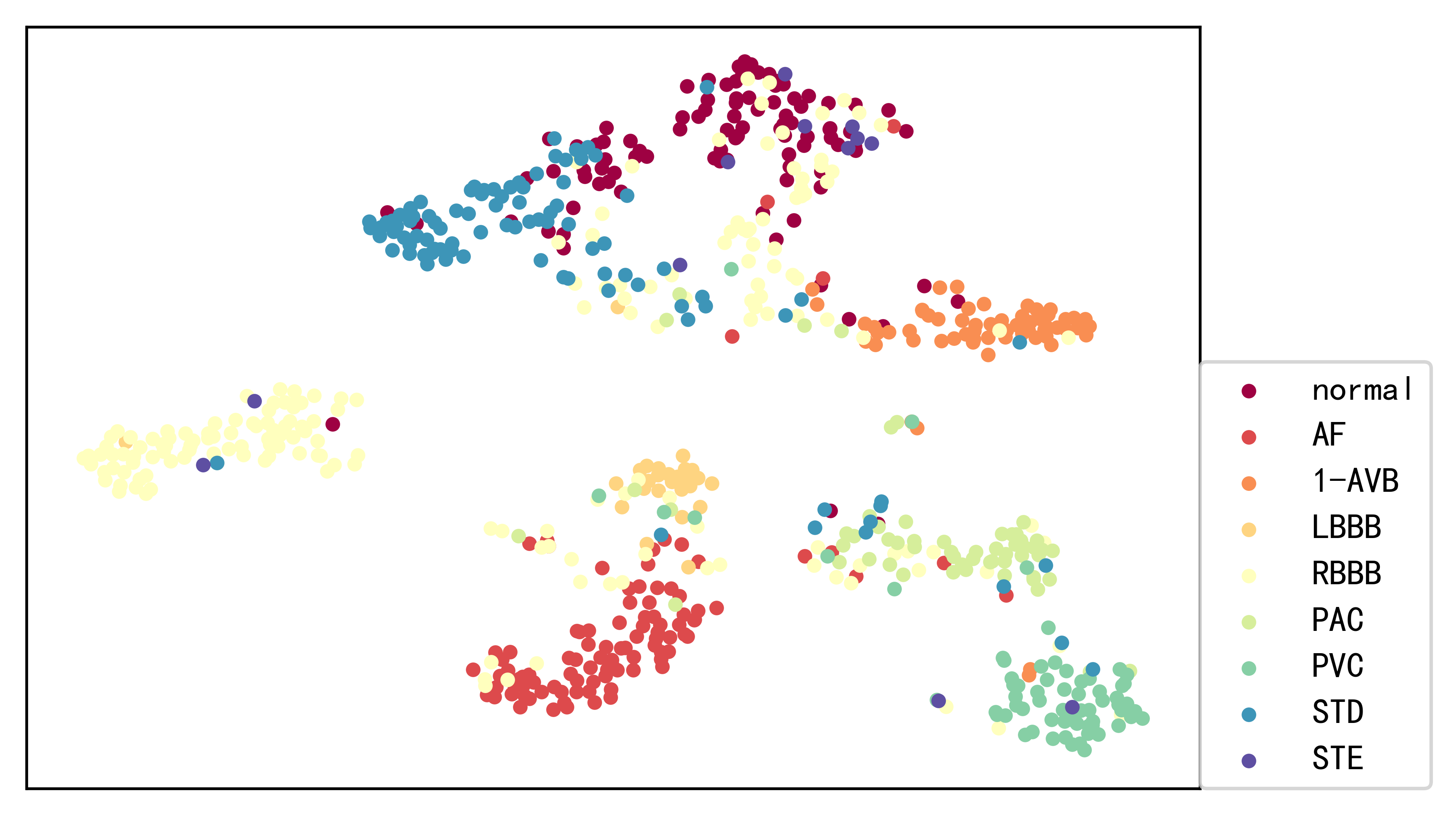}}
\subfloat[branch for lead III]{
\label{III}
\includegraphics[width=0.29\textwidth]{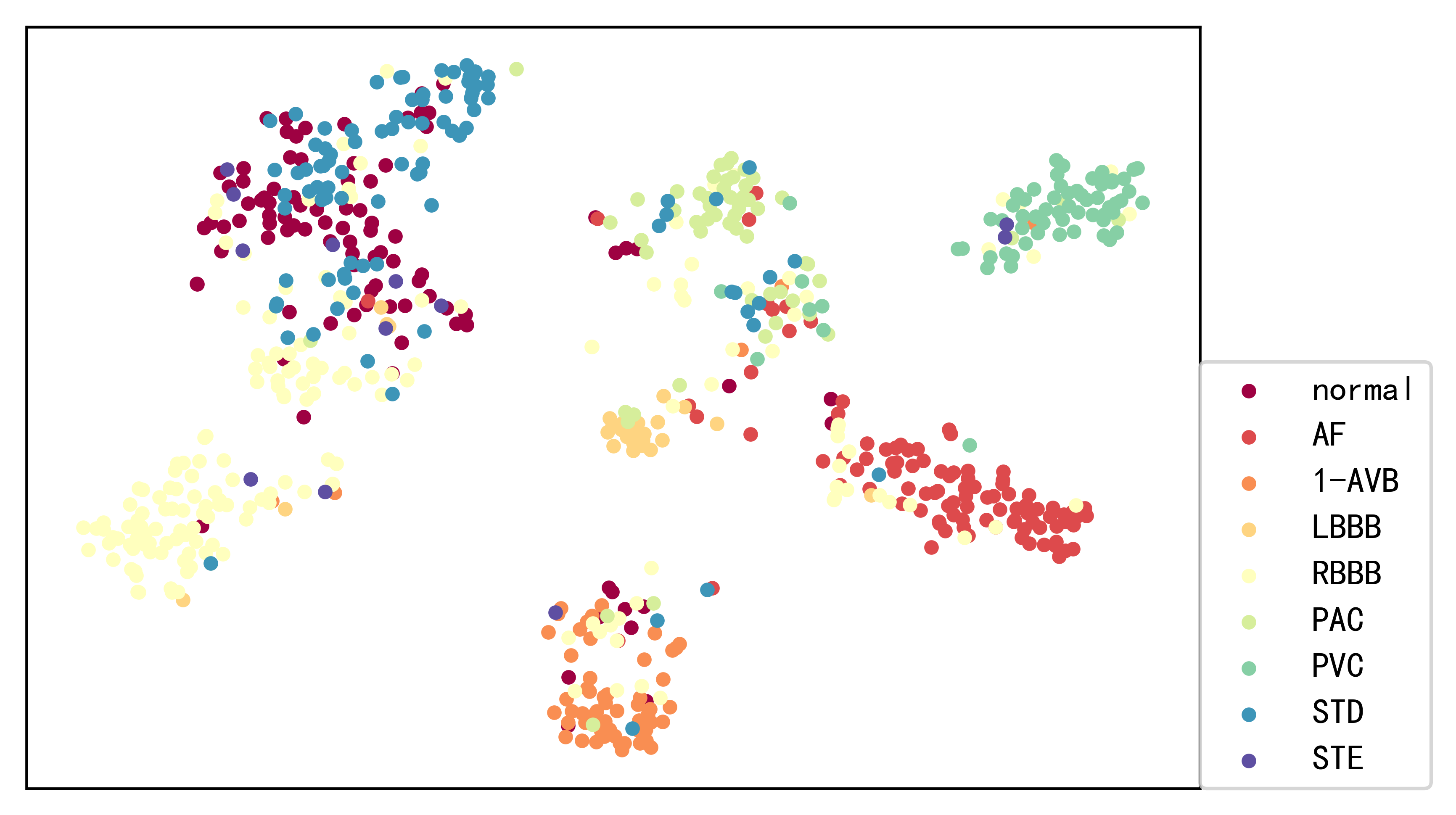}} \\
\subfloat[branch for lead avR]{
\label{avR}
\includegraphics[width=0.29\textwidth]{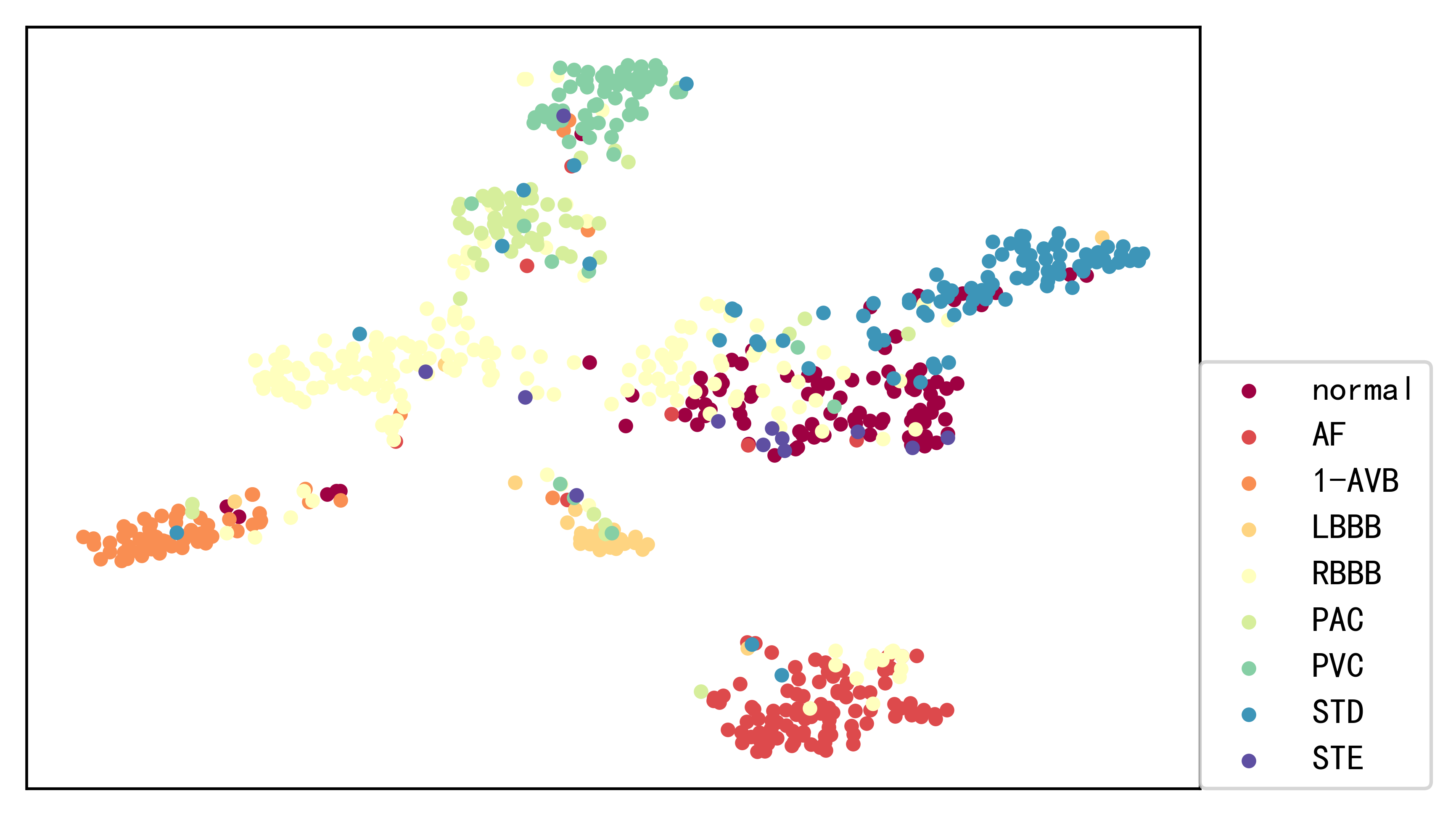}}
\subfloat[branch for lead avL]{
\label{avL}
\includegraphics[width=0.29\textwidth]{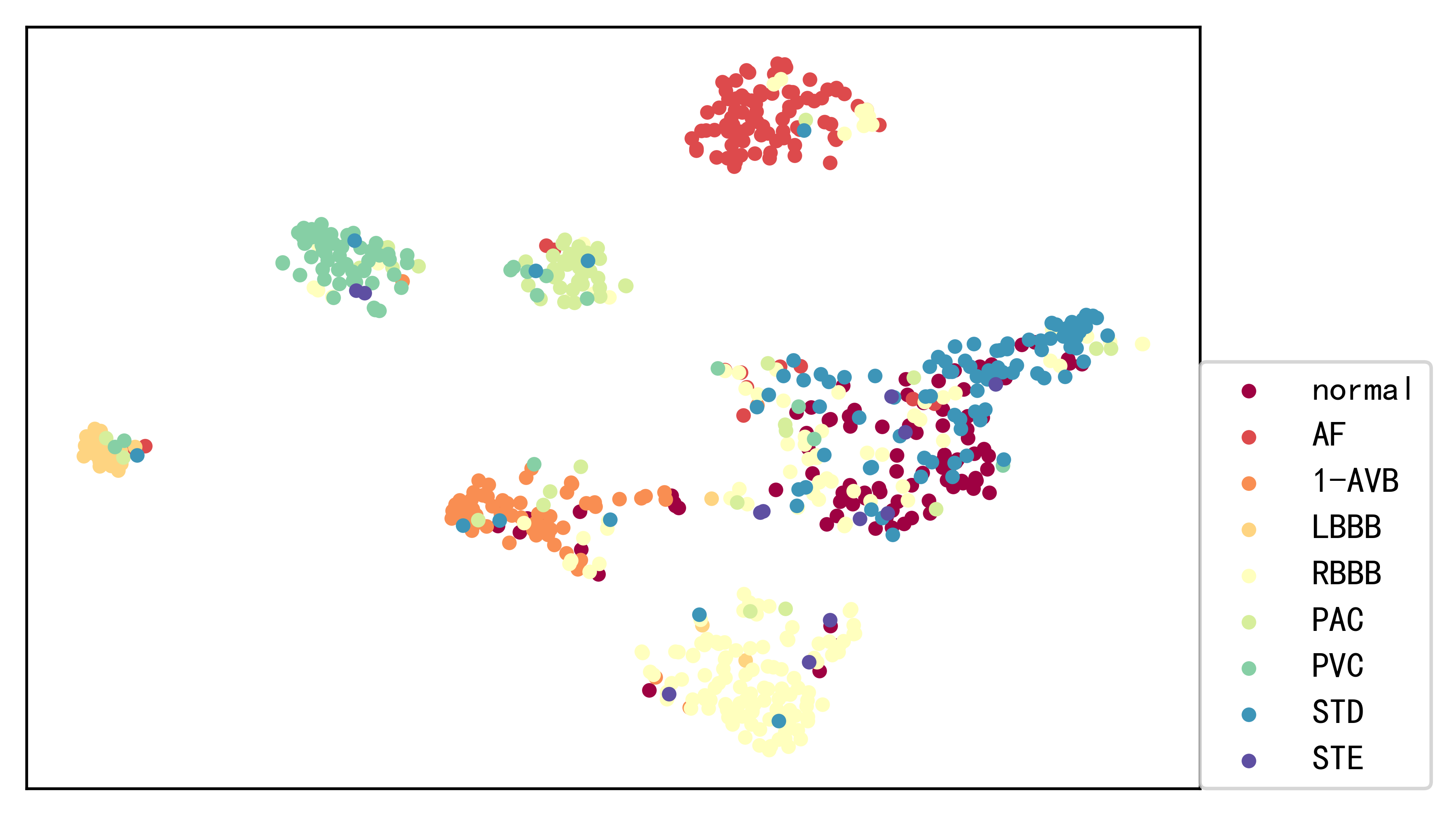}}
\subfloat[branch for lead avF]{
\label{avF}
\includegraphics[width=0.29\textwidth]{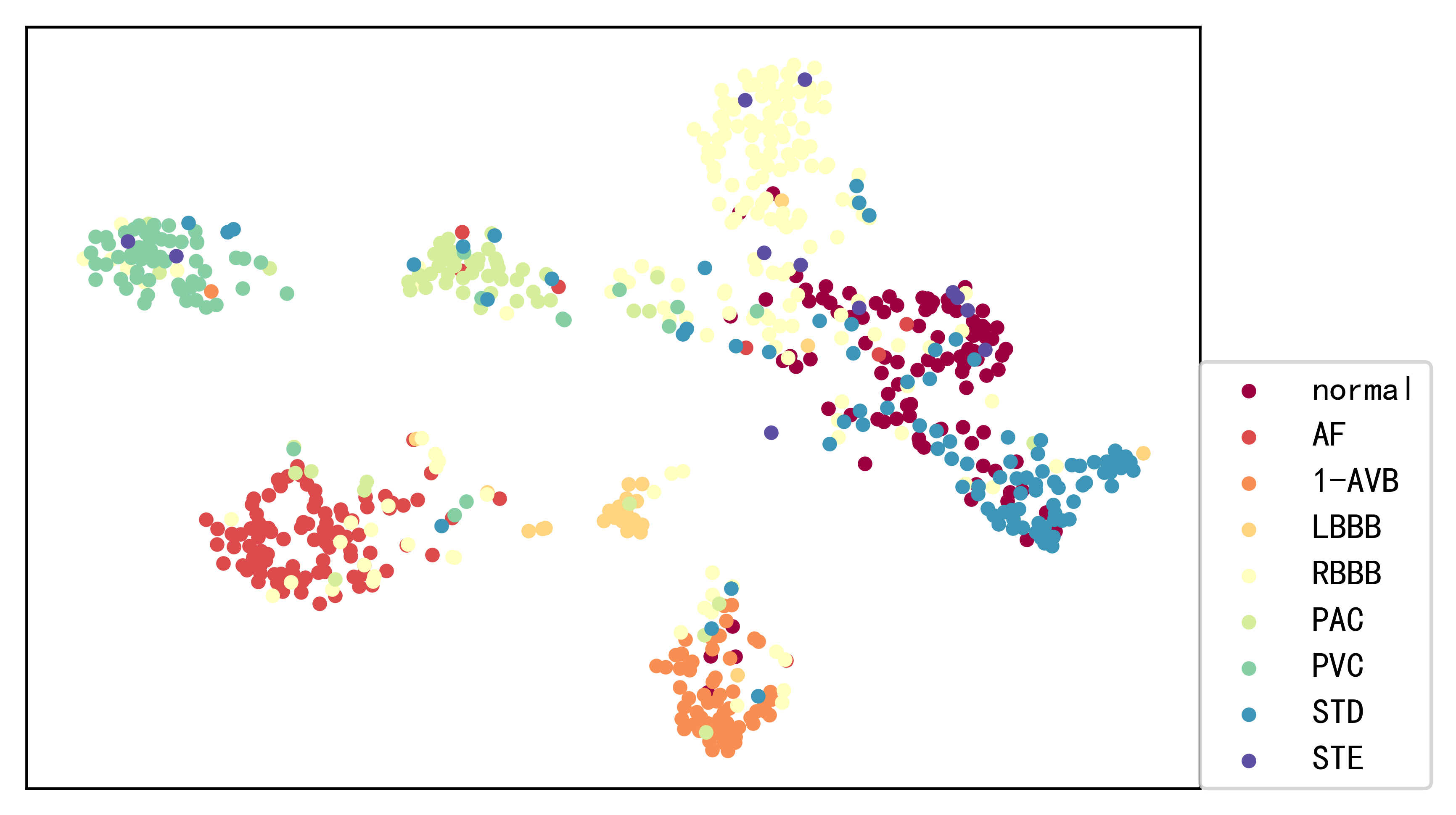}} \\
\subfloat[branch for lead V1]{
\label{V1}
\includegraphics[width=0.29\textwidth]{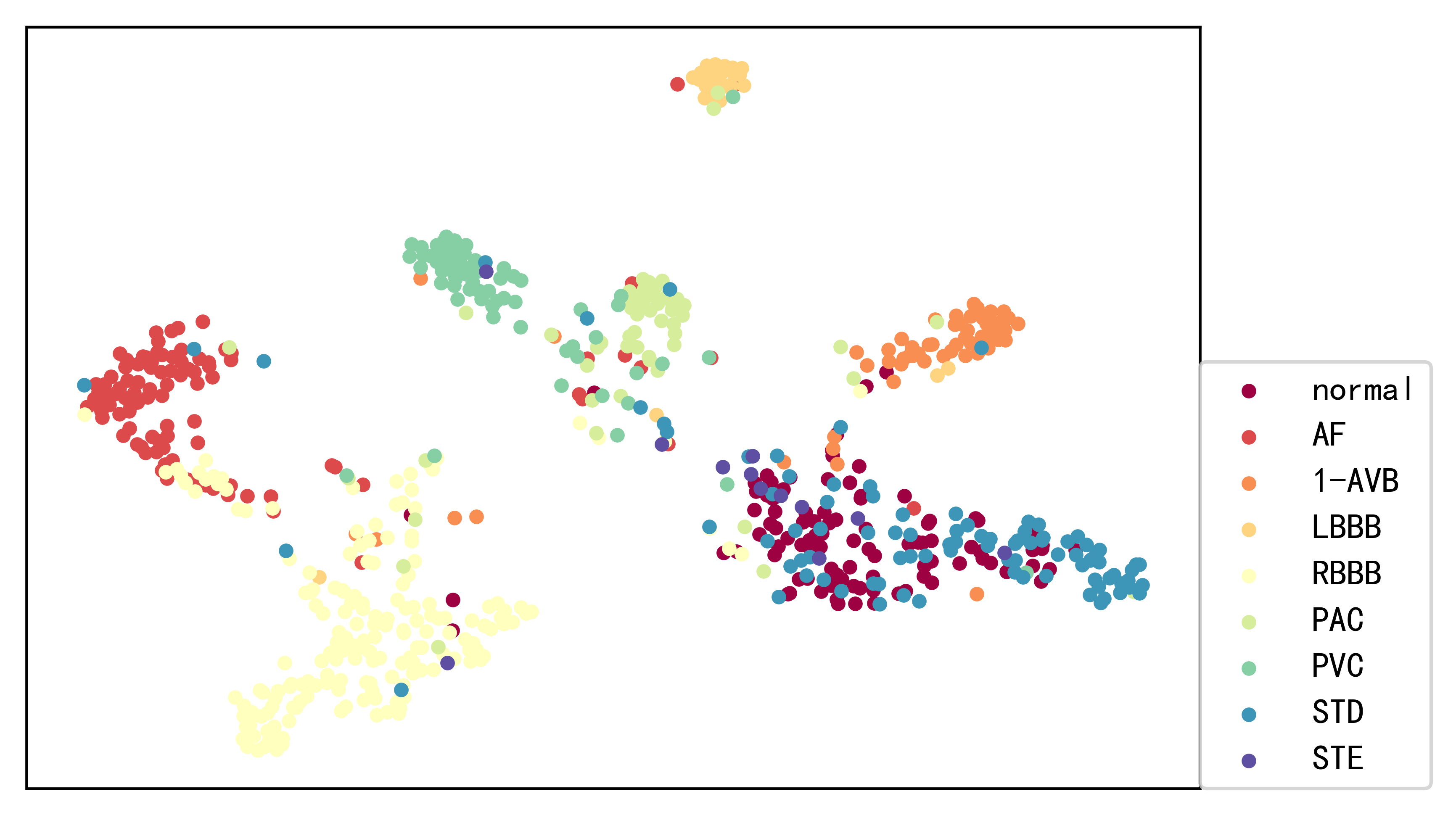}}
\subfloat[branch for lead V2]{
\label{V2}
\includegraphics[width=0.29\textwidth]{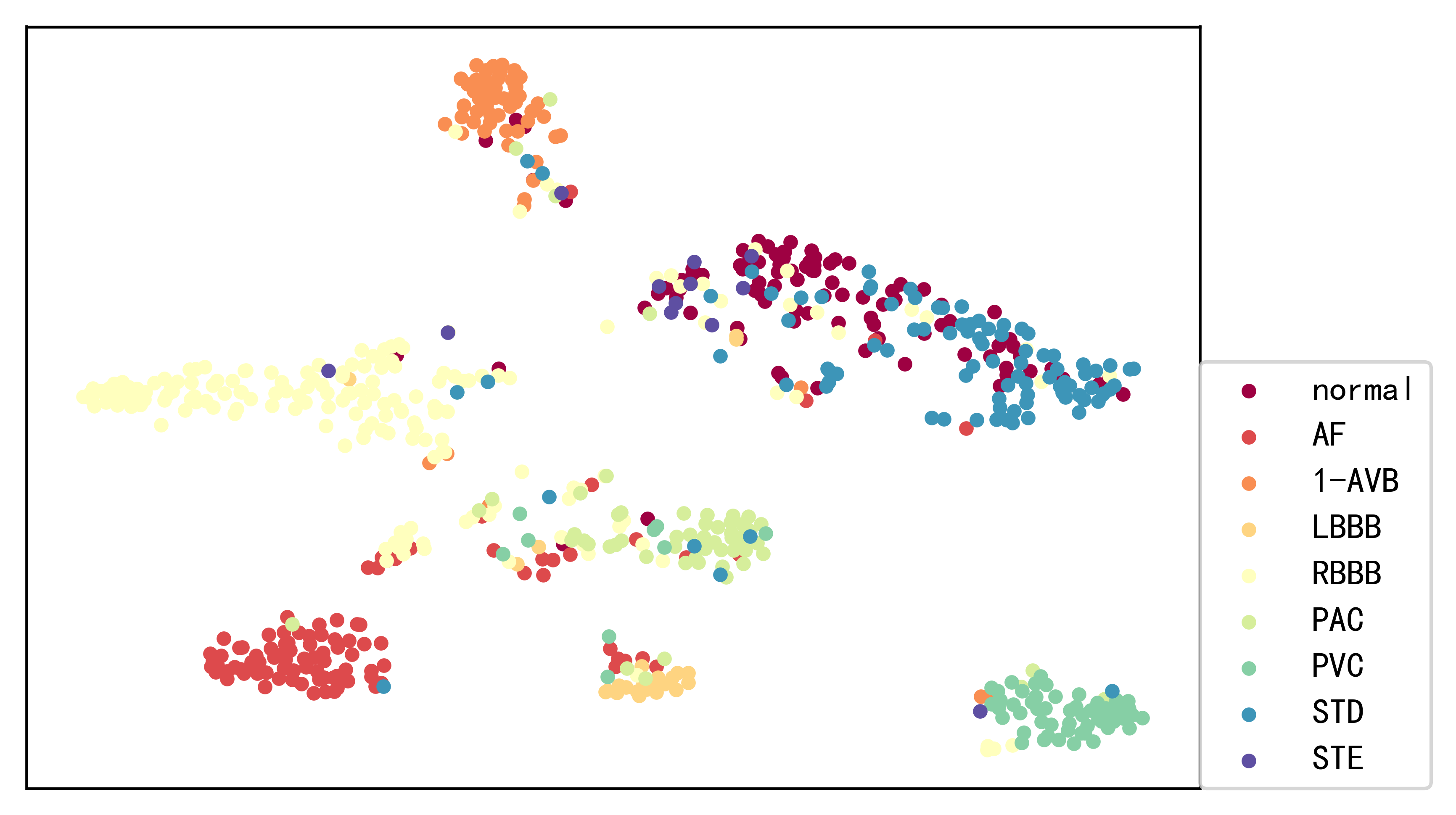}}
\subfloat[branch for lead V3]{
\label{V3}
\includegraphics[width=0.29\textwidth]{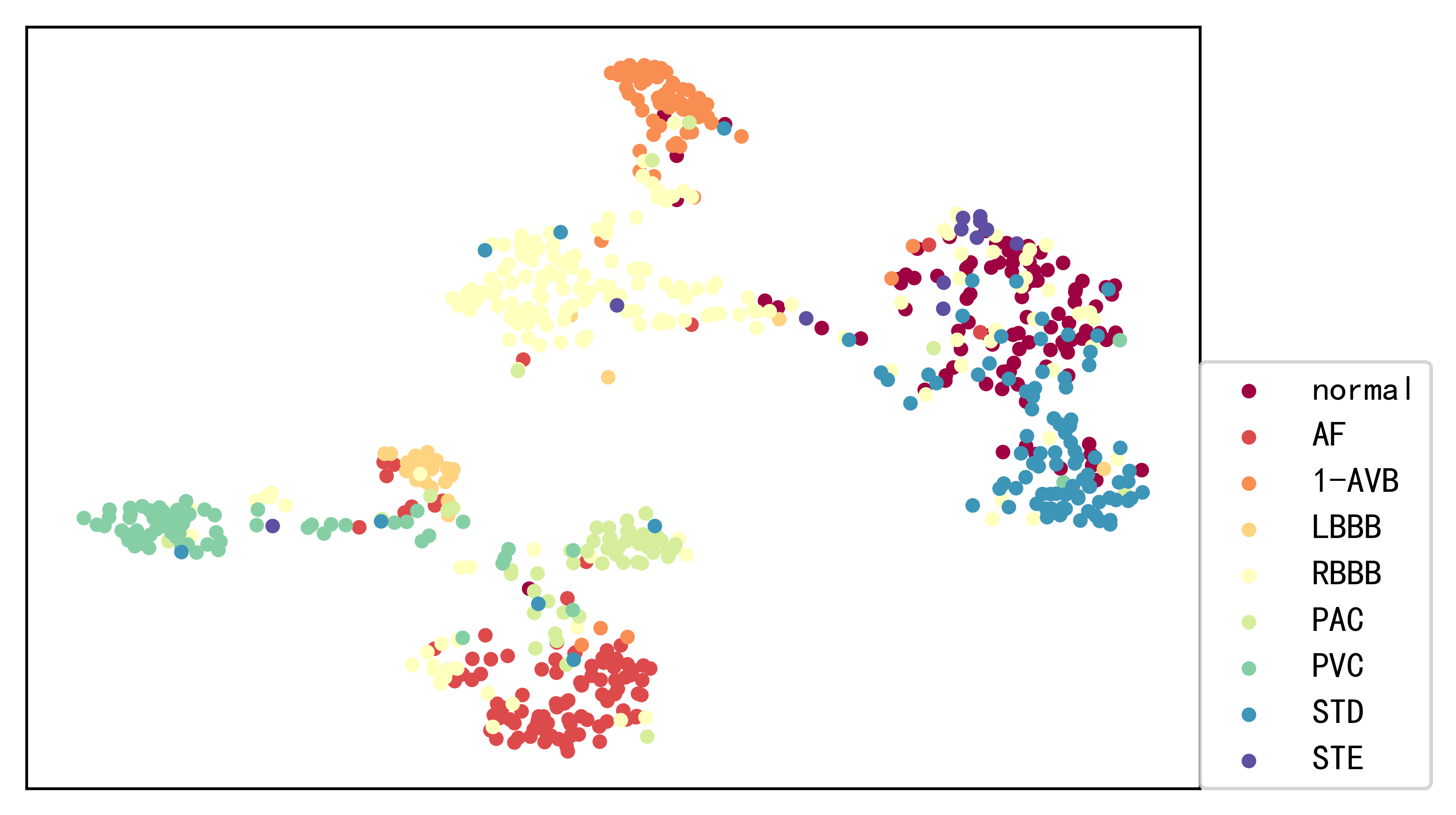}} \\
\subfloat[branch for lead V4]{
\label{V4}
\includegraphics[width=0.29\textwidth]{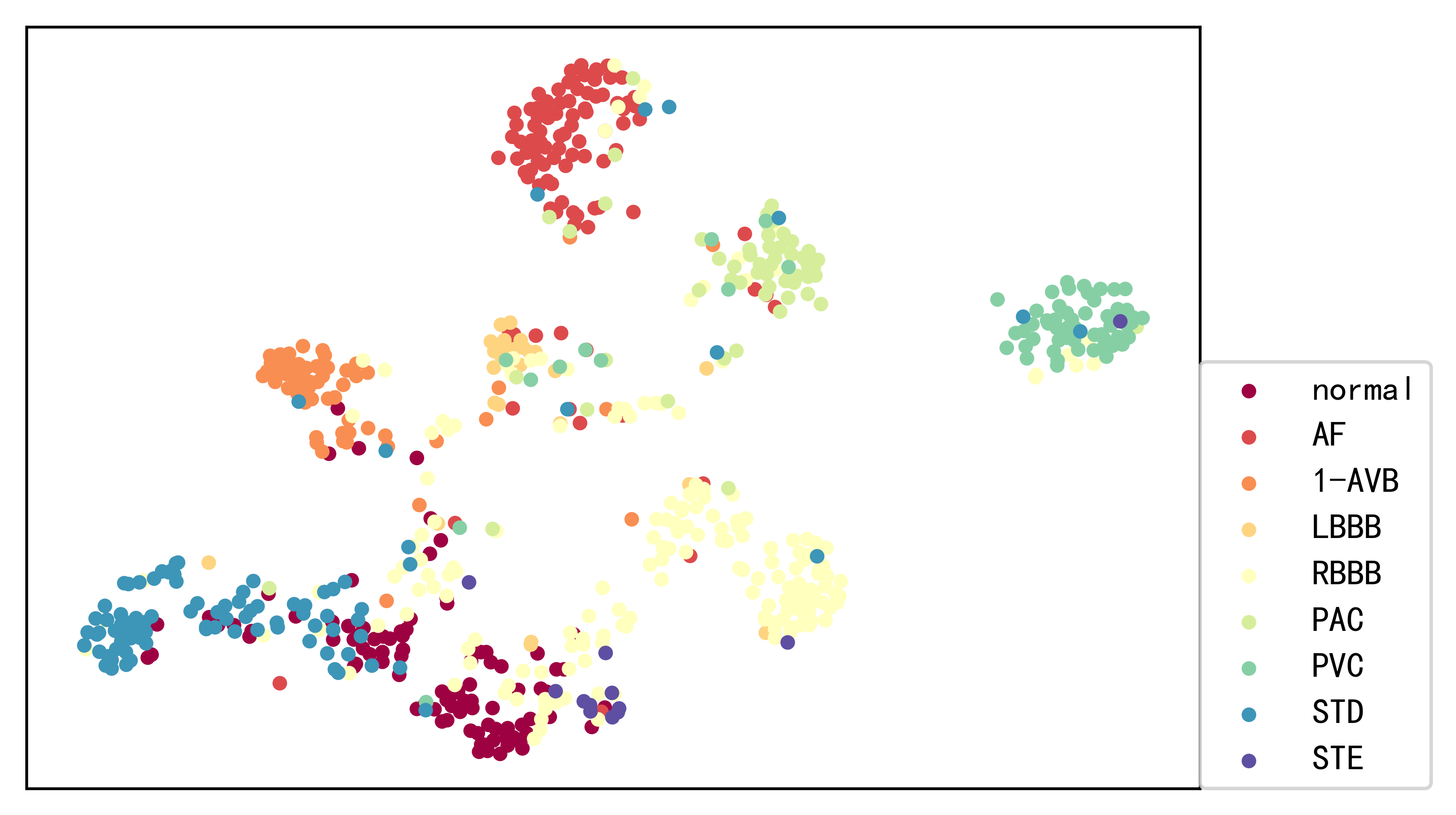}}
\subfloat[branch for lead V5]{
\label{V5}
\includegraphics[width=0.29\textwidth]{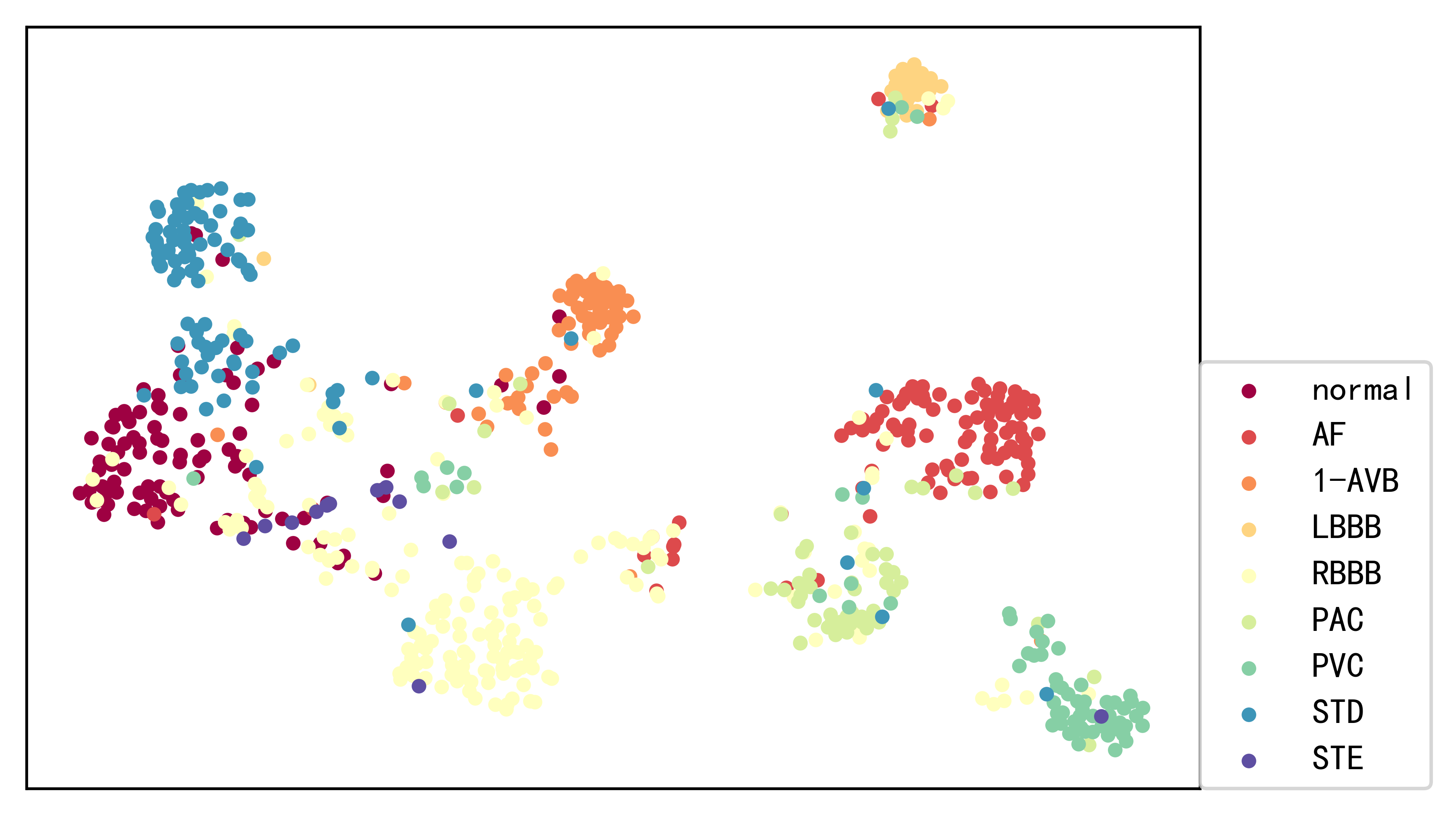}}
\subfloat[branch for lead V6]{
\label{V6}
\includegraphics[width=0.29\textwidth]{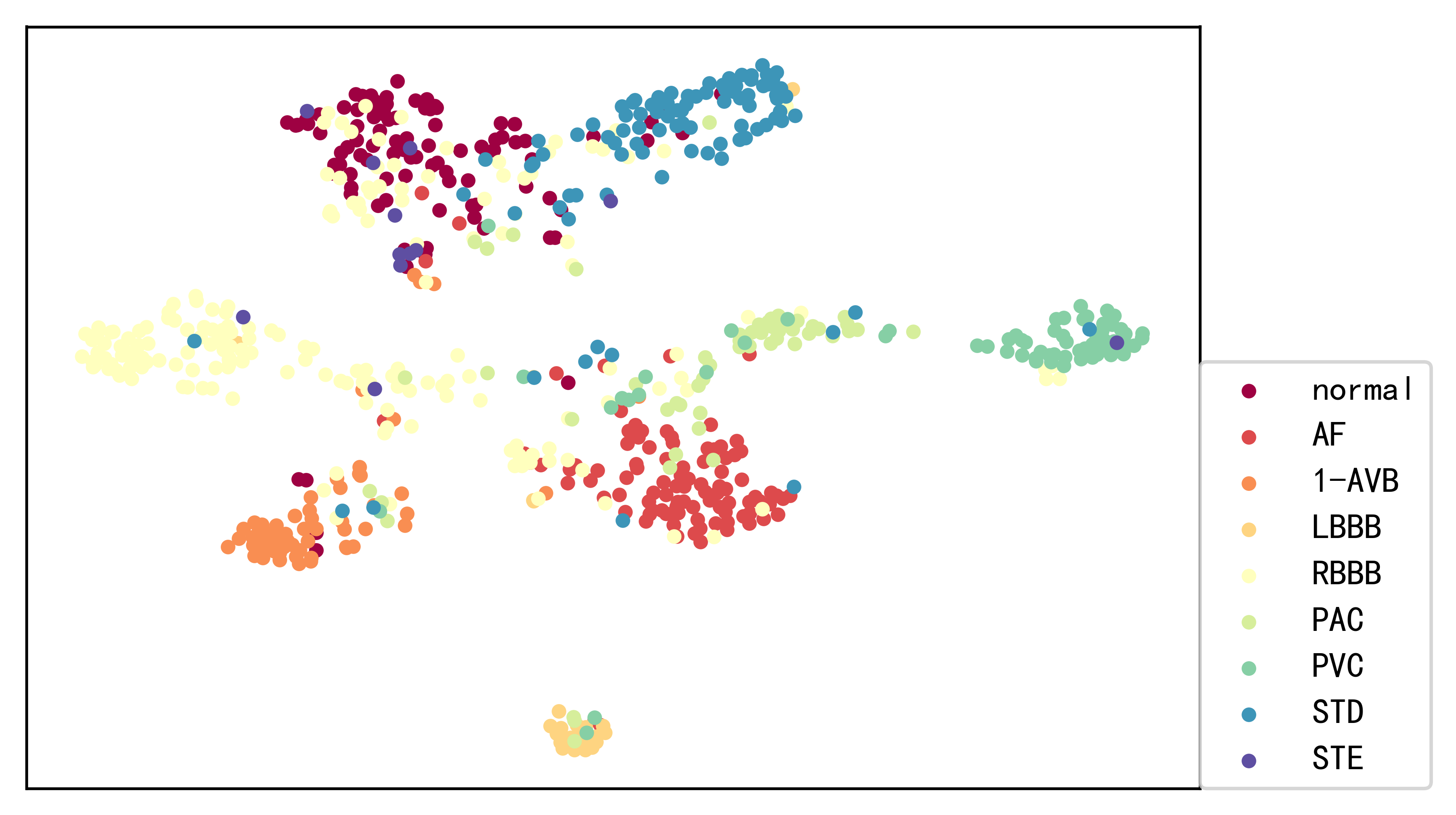}} \\
\caption{The t-SNE visualization for the performance of MLBF-Net (a) and each branch (b-m).}
\label{visualization}
\end{figure*}
\subsection{Ablation Studies}
To verify the effectiveness of main components in MLBF-Net, we also evaluated several variants of the network. The following experiments involved in the Discussion section were evaluated through 10-fold cross validation, not the private test set. Unlike the experiment conducted in the Experiment section, the public dataset was randomly divided into ten subsets, with each subset taking turns as the test set. The remaining records were further divided into training data and validation data, of which 11.12\% was as validation data. Finally, the ratio of training data, validation data, and test data in each fold is 8: 1: 1.

\subsubsection{Single-Branch Framework vs. Multi-Branch Framework}
When dealing with multi-lead ECG signals, the regular frameworks are that multi-lead ECG is concatenated into an integrated matrix and then input into a feature extractor or deep neural network. They can be regarded as single-branch. Despite the simplicity of such network frameworks, the diversity of multi-lead ECG signal is neglected. To analyze the effect of multi-lead diversity, we take the BranchNet to classify 12-lead ECG signals. Table \ref{table3} shows the performances of BranchNet with a single branch against MLBF-Net with multiple branches. In terms of network architecture, BranchNet is a branch of MLBF-Net, but it should be noted that in this experiment, the input to BranchNet is 12-lead ECG signals, not single-lead ones. We can observe that multi-branch model is far superior to single-branch model for all classes in the $F_1$ score, gaining 3.7\% increase in the average $F_1$ score. It is proved that lead-specific features do considerably improve the classification performance of MLBF-Net.

\subsubsection{Single-Loss Optimization vs. Multi-Loss Optimization}
\begin{table}[!t]
\centering
  \caption{\label{table4}Classification performance comparison for single-loss and multi-loss optimization by 10-fold cross validation (mean$\pm$SD).}
  \begin{tabular}{ccccccccccc}
    \toprule
    Setting & average $F_1$ \\
    \midrule
    single loss & 0.828$\pm$0.021 \\
    $\lambda$=0.1 & 0.837$\pm$0.020 \\
    $\lambda$=0.2 & 0.844$\pm$0.017 \\
    $\lambda$=0.5 & 0.849$\pm$0.017 \\
    \textbf{$\lambda$=1} & \textbf{0.852$\pm$0.021} \\
    $\lambda$=2 & 0.843$\pm$0.014 \\
    $\lambda$=5 & 0.848$\pm$0.023 \\
    $\lambda$=10 & 0.847$\pm$0.023 \\
    \bottomrule
  \end{tabular}
\end{table}

The hyperparameter $\lambda$ governs the participation rate of lead-specific features in our model. Here, we conducted some experiments to explore the sensitiveness of this hyperparameter. The first experiment was implemented for evaluating the effectiveness of the multi-loss co-optimization strategy, in which the attention modules from all branches are removed. These attention modules are used to generate lead-specific features that correspond to all branch losses. Finally, we only keep the loss for the concatenated network. Table \ref{table4} shows the above experimental result. It is observed that multi-loss co-optimization improves the $F_1$ score by 2.4\% (0.828-0.852). Furthermore, extra experiments were conducted to analyze the effect of different participate rates of lead-specific features by setting $\lambda$ from 0 to 10 in our model. As we can see from Table \ref{table4}, the classification performance is highest on the condition that lead-specific features with diversity and comprehensive features with integrity are equally important ($\lambda$=1). When either of lead-specific features and comprehensive features is more predominant, the classification performance is slightly worse than that of their equal participation rate. It is clear that the arrhythmia detection performance is significantly improved through multi-loss co-optimization to jointly learning diversity and integrity.

\subsection{Visualization of Learned Features}
The t-distributed stochastic neighbor embedding (t-SNE) \cite{dermaaten2008visualizing} visualizes high dimensional data in a two or three-dimensional map. Here, we introduced t-SNE algorithm to evaluate the proposed method visually. The attention modules in each branch and the concatenated network output the 24-dimensional features. These features are as the input of t-SNE to visualize the performance of learned the comprehensive features and lead-specific features, shown in \ref{multi-lead} and Fig. \ref{I}-\ref{V6}. As we can see from Fig. \ref{I}-\ref{V6}, the dots of different colors represent the extracted features of different types of ECG signals with varying degrees of overlap, meaning that only lead-specific features of a single lead are incapable of distinguishing multi-class arrhythmias well. Due to the small number of STE samples, the purple dots representing STE are difficult to observe. Excluding STE, the heaviest overlapping is in distinguishing STD from Normal signals, probably because the tiny changes in ST-segment are prone to be contaminated by various noises. In Fig. \ref{multi-lead}, the colorful dots are separated obviously. Thus, the features learned by multi-branch with multi-loss co-optimization are discriminative for classifying 9 ECG classes. It can be observed from this figure that the yellow dots representing RBBB have two clusters, possibly because the used dataset contains both complete RBBB and incomplete RBBB in which QRS complex durations are different.

\subsection{Model Parameter}
In this subsection, the parameter amount of the proposed MLBF-Net is analyzed, shown in Table. \ref{parameters}. The multi-branch architecture seems to introduce a large number of parameters exponentially. In fact, our model is still lightweight in comparison with many previous deep learning-based studies for arrhythmia detection. In \cite{hannun2019cardiologist}, the comparable performance to cardiologists was reported for the identification of 12 types of arrhythmias by using a 34-layer ResNet. However, the training parameters were as high as about 10.47 million. Yao et al. \cite{yao2020multi} evaluated their proposed method on the same independent test set as ours, and obtained a $F_1$ score of 0.812. In their study, the parameters were about 4.98 million. As calculated in Table. \ref{parameters}, only 0.42 million parameters need to be trained in our model, realizing a tens of times parameter reduction than \cite{hannun2019cardiologist} and \cite{yao2020multi}.

\begin{table}[!]
  \centering
  \caption{\label{parameters}Parameter amount for the proposed model, MLBF-Net.}
  \begin{tabular}{l}
    \centering
	\includegraphics[width=0.95\linewidth]{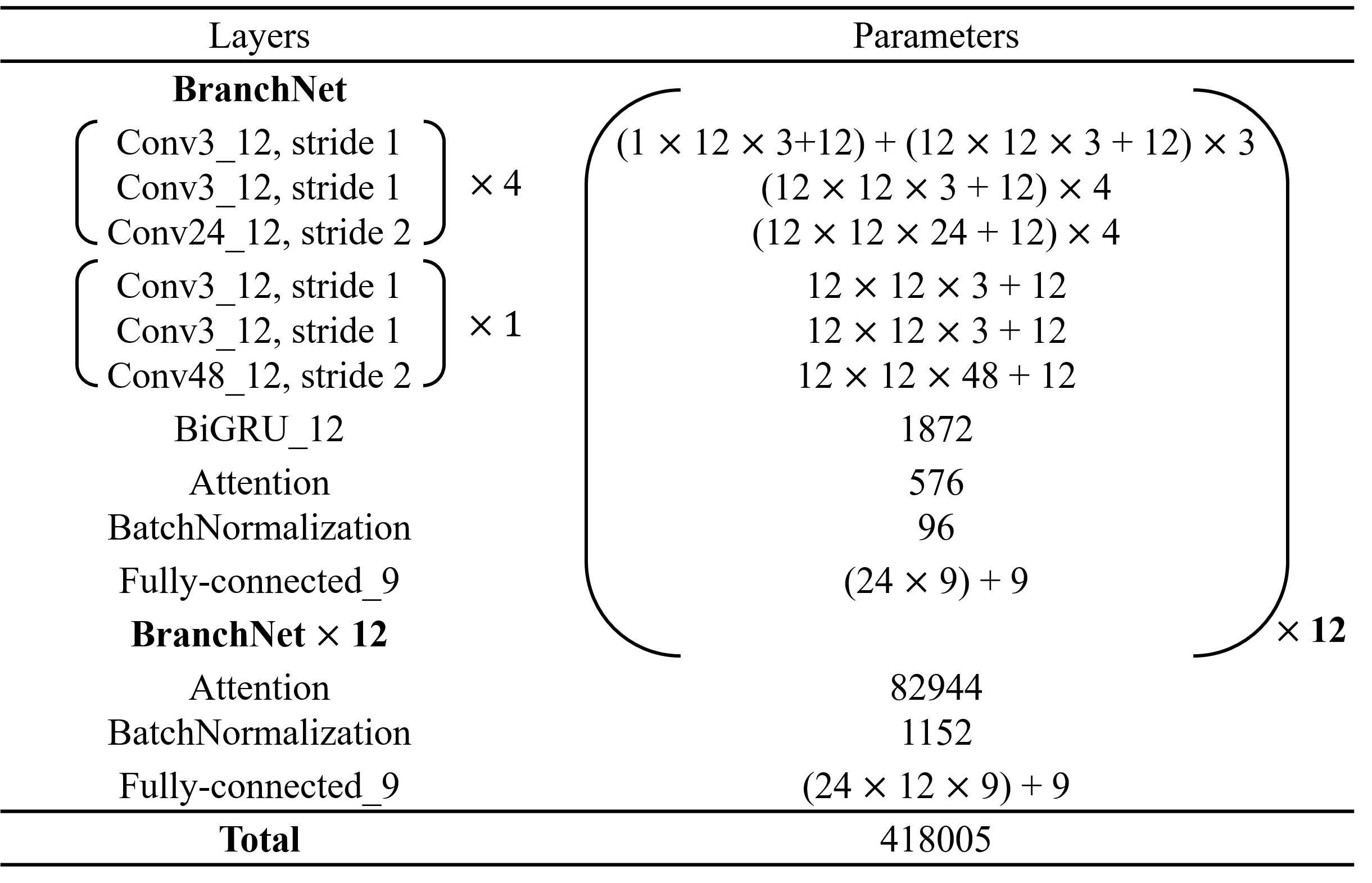}
  \end{tabular}
\end{table}

\section{Conclusion}
In this paper, we propose a novel end-to-end Multi-Lead-Branch Fusion Network (MLBF-Net) for ECG classification using 12-lead ECG records. MLBF-Net fully utilizes the diversity and integrity of multi-lead ECG by integrating multiple losses to optimize lead-specific features and comprehensive the 12-lead features collaboratively. We demonstrate that our MLBF-Net reaches the highest arrhythmia classification performance on China Physiological Signal Challenge 2018 which is an open 12-lead ECG dataset. In addition, compared with many existing deep neural networks, MLBF-Net is a parameter-efficient model that is less prone to overfitting, despite its multiple branches architecture. The proposed model has the advantages of both high screening capability and light weight. Therefore, it has the potential to be applied in clinical applications and daily monitoring.

\bibliography{ref}

\end{document}